\documentclass[12pt]{article}
\usepackage{amsmath}
\usepackage{graphicx}
\usepackage{natbib}
\usepackage{amssymb}
\usepackage{algorithm}
\usepackage{multirow}
\usepackage[noend]{algpseudocode}

\usepackage{url} 

\newcommand{\blind}{0}

\newcommand{\Y}{{\mbox{\boldmath $Y$}}}

\newcommand{\mbf}[1]{\mbox{\boldmath${#1}$}}
\newcommand{\X}{{\mbox{\boldmath $X$}}}

\newcommand{\Z}{{\mbox{\boldmath $Z$}}}

\newcommand{\W}{{\mbox{\boldmath $W$}}}

\newcommand{\D}{{\mathcal{D}}}

\newcommand{\be}{\mbf{\beta}}

\newcommand{\bgamma}{{\mbox{\boldmath $\gamma$}}}

\newcommand{\bTheta}{{\mbox{\boldmath $\Theta$}}}

\newcommand{\bfeta}{{\mbox{\boldmath $\xi$}}}

\newcommand{\bvarphi}{{\mbox{\boldmath $\bvarphi$}}}

\newcommand{\beast}{\mbf{\beta}^\ast}

\begin{document}

\def\spacingset#1{\renewcommand{\baselinestretch}%
{#1}\small\normalsize} \spacingset{1}


\if0\blind
{
   \title{\bf Efficient Sparse High-dimensional Linear Regression with a Partitioned Empirical Bayes ECM Algorithm}
  \author{Alexander C. McLain$^a$, Anja Zgodic$^a$, and Howard Bondell$^b$ \vspace{0.2cm}\\ 
    $^a$Department of Epidemiology and Biostatistics, \\ University of South Carolina \vspace{0.2cm}\\   $^b$School of Mathematics and Statistics \\ University of Melbourne}
  \maketitle
} \fi

\if1\blind
{
  \bigskip
  \bigskip
  \bigskip
  \begin{center}
    {\LARGE\bf Efficient Sparse High-dimensional Linear Regression with a Partitioned Empirical Bayes ECM Algorithm}
\end{center}
  \medskip
} \fi

\bigskip
\begin{abstract}
Bayesian variable selection methods are powerful techniques for fitting and inferring on sparse high-dimensional linear regression models. However, many are computationally intensive or require restrictive prior distributions on model parameters. In this paper, we proposed a computationally efficient and powerful Bayesian approach for sparse high-dimensional linear regression. Minimal prior assumptions on the parameters are required through the use of plug-in empirical Bayes estimates of hyperparameters. Efficient maximum \textit{a posteriori} (MAP) estimation is completed through a Parameter-Expanded Expectation-Conditional-Maximization (PX-ECM) algorithm. The PX-ECM results in a robust computationally efficient coordinate-wise optimization which -- when updating the coefficient for a particular predictor -- adjusts for the impact of other predictor variables. The completion of the E-step uses an approach motivated by the popular two-group approach to multiple testing. The result is a PaRtitiOned empirical Bayes Ecm (PROBE) algorithm applied to sparse high-dimensional linear regression, which can be completed using one-at-a-time or all-at-once type optimization. We compare the empirical properties of PROBE to comparable approaches with numerous simulation studies and analyses of cancer cell drug responses. The proposed approach is implemented in the R package \texttt{probe}. 
\end{abstract}

\noindent%
\vfill

\newpage
\spacingset{1.75} 
\section{Introduction}
\label{sec:intro}

A common goal in the analysis of high-dimensional data is to use a set of $n$ replicates of $M$ predictors ($\X$) hypothesized to have a relationship with a continuous outcome of interest ($\Y$) to develop a linear model that can estimate regression coefficients, and predict a future outcome given new predictor data. In applications to data from genetics, neuroimaging, financial forecasting, or text mining (among others) $M \gg n$ scenarios can easily arise. In this setting, some additional assumptions on the linear model as required, for example, the sparsity assumption, which posits that most of the predictors will have zero impact on the outcome. 

Bayesian variable selection procedures are powerful methods for estimating the properties of regression coefficients and identifying which variables have a meaningful impact on the outcome. Further, they can incorporate high-dimensional predictor variables and sparsity through prior distributions on the regression coefficients. Many prior distributions use a spike-and-slab form, which incorporates binary latent variables that indicate when a predictor has a non-zero or ``large'' impact on $\Y$ \citep{MitBea88, GeoMcC93, Casetal15}. Other choices of priors often take the form of scale mixtures of Gaussians, for example, the Horseshoe \citep{Caretal10} or the Bayesian bridge \citep{Poletal14}, among others. After a prior distribution is specified, estimation of the posterior distribution or the maximum \textit{a posteriori} (MAP) values can proceed. While Markov chain Monte Carlo \citep[MCMC, ][]{Lietal08,BonRei12,Chaeetal19} is technically an option when $M \gg n$, sampling over the high-dimensional parameter space can be computationally prohibitive with chains that can be slow to converge. This has led to the use of more computationally feasible estimation methods, such the expectation maximization (EM) algorithm \citep{RocGeo14,Roc18,RocGeo18}, empirical Bayes \citep{GeoFos00,Maretal17,MarTan20}, and Variational Bayes \citep[VB,][]{CarSte12,Bleetal17,Wanetal20,RaySza22}.

The main challenges that arise in Bayesian variable selection are choosing appropriate prior distributions and computational complexity. Regarding the former, the results can be sensitive to the choice of the prior distribution, particularly in the high-dimensional setting \citep{Casetal15,MarTan20}. Using uninformative priors, which are designed to have a small impact on the results, is one option for avoiding these issues \citep{MitBea88}. When a spike-and-slab prior is used, it is not required that the `slab' portion of the prior penalizes or shrinks the coefficients since the `spike' portion can absorb small non-important effects. However, using Bayes factors to compare null models (e.g., without predictors) to alternative models (with some predictors) will favor the null model, regardless of the data, when flat uninformative priors are used. This is known as \emph{Barlett's paradox} \citep[][]{Bar57,KloLem70} and (in the limit) results in Bayes factors placing all posterior probability on the submodel that omits all predictors. This has led to a departure from uninformative priors in favor of parametric options, which -- while being able to use Bayes Factors -- can be sensitive to the parametric specification.

In this paper, we perform Bayesian variable selection with an uninformative spike-and-slab prior on the regression parameters, which has not yet been used in the high-dimensional setting. We focus on MAP estimation of regression parameters and the residual variance, along with the posterior expectation of latent variable selection indicators. We use a quasi Parameter-Expanded Expectation-Conditional-Maximization (PX-ECM), which is a combination of the ECM \citep{MenRub92,MenRub93,vanetal95} and PX-EM \citep{Liuetal98}. With the standard EM, the M-step would require optimization with respect to the high-dimensional regression parameter, which is not feasible with uninformative priors. The ECM breaks the M-step into computationally simple coordinate-wise (predictor level) optimizations. Within the optimization for a given predictor, the PX portion adds a parameter that scales the impact of the remaining predictor variables. Unlike other coordinate-wise optimization approaches, this does not assume that the impact of remaining predictor variables is fixed. The benefits of this formulation are that the impact of the remaining predictor variables only needs to be known up to a multiplicative constant, we can account for the increase in uncertainty due to the dependence between a given predictor and the impact of the remaining predictor variables, and it adds stability to the algorithm. 

Standard computation of the E-step consists of using Bayes factors to estimate the probability of the latent variable selection indicators. This is not an option in our setting since Bayes factors will favor the null model with uninformative priors. Here, instead of changing the prior to accommodate the use of the Bayes factors, we accommodate the prior by using plug-in empirical Bayes estimates to complete the E-step.  Since the seminal work of \cite{GeoFos00}, empirical Bayes has been thoroughly studied and applied to various variable selection models \citep{ScoBer10,Petetal14,BelGho20,MarTan20,Kimetal22}. In this paper, we take an approach that is motivated by the popular two-groups approach to multiple testing \citep{Efron2001,Sun2007}, with empirical Bayes estimates of the hyperparameters. The empirical Bayes procedure requires estimates of the posterior variance of the regression parameters, which are obtained using Taylor-expansion of the complete data posterior variance.

We refer to the main components of our optimization algorithm as a PaRtitiOned empirical Bayes Ecm (PROBE) algorithm. We develop \textit{one-at-a-time} and \textit{all-at-once} variants of the PROBE algorithm, which are analogous to least squares estimation with the Gauss-Seidel and Jacobi methods, respectively \citep{Mas95,LevLew10}. In each iteration of the one-at-a-time version, parameter values are sequentially updated (i.e., maximized with respect to), with each update accounting for all previous updates in the current iteration. In contrast, the updates in the all-at-once variant only account for the updates from the previous iteration. 

The paper is outlined as follows. In Section \ref{sec.methods}, we introduce and motivate our modeling framework and show how the E- and M-steps are performed. In Section \ref{sec.estimation}, we give full details on the implementation of the one-at-a-time and all-at-once variants of the PROBE algorithm. In Section \ref{sec.sim}, we evaluate the predictive ability of the proposed approach via simulation and compare it with common approaches to high-dimensional linear regression. In Section \ref{sec.anal}, we use the proposed method to analyze drug response in cancer cell data and compare the properties of the estimates to other approaches. Section \ref{sec.disc} provides a discussion of our method and future research directions.


\section{Methods}\label{sec.methods}

Let $X_{im}$ denote predictor $m$ for observation $i$, $m=1,2,\ldots,M$ and $i=1,2,\ldots,n$ with $\X_m=(X_{1m},\ldots,X_{nm})'$ an $n \times 1$ vector, $\X=(\X_1, \ldots , \X_M)$ an $n \times M$ matrix, and $Y_i$ the outcome of interest for observation $i$ with $\mbf{Y}=(Y_{1},Y_{2},\ldots,Y_{n})'$. For ease of presentation, we assume that $\Y$ and $\X$ have been centered. Further, we assume that all variables in $\X$ are being subjected to the sparsity assumption. The linear model is given by
\begin{equation}\label{eq.model}
\Y = \X (\bgamma \circ \be)  +   \mbf{\epsilon},
\end{equation}
where $\bgamma\circ\be$ is a Hadamard product of $\be \in \mathbb{R}^M$ and $\bgamma \in \{0,1\}^M$, and $\epsilon_{i}$ are independent with $Var(\epsilon_{i}) = \sigma^2$. We assume that there exists true $\be$, $\bgamma$, and $\sigma^2$ which are fixed and unknown. If $\gamma_m=0$ then we use the convention that $\beta_m=0$. To estimate these quantities, we use a Bayesian framework with priors
\begin{eqnarray*}
p(\be ) &=& \prod_{m=1}^M f_\beta(\beta_m), \\
f_\beta &\propto& 1 \\
p(\bgamma|\pi) &=& \pi^{M_1} (1-\pi)^{M-M_1},\\
\sigma^2  &\sim& IG\left(a, b \right),
\end{eqnarray*}
and $\pi = \Pr(\gamma=1) \sim f_\pi$ where $M_1 = \sum_m \gamma_m$ and $IG$ denotes an inverse-Gamma distribution. Throughout, we use an improper $IG$ prior on $\sigma^2$ with $(a,b)=\left(-\frac{3}{2},0\right)$ and, for brevity, we do not explicitly condition these parameters in the probability functions below. We leave the prior for $\pi$ unspecified and describe how it will be treated in Section \ref{sec.e.step}. Let $\bTheta = (\sigma, \mbf{\beta}, \pi)$ denote the full parameter vector, $\mathcal{D} = (\Y,\X)$ denote the observed data, $\mbf{A}_{/m}$ denote the matrix, vector, or set $\mbf{A}$ without column or element $m$, and (for brevity) $\mbf{a}\mbf{b} \equiv \mbf{a}\circ \mbf{b}$ for all vectors $\mbf{a}$ and $\mbf{b}$. 

A common approach to fitting such models is to use the EM algorithm to find the values of $\bTheta$ that maximize the posterior distribution, i.e., the MAP estimates. The posterior distribution of $\bTheta$ given $\bgamma$ is
\begin{equation}\label{eq.Likelihood}
p(\bTheta|\mathcal{D},\bgamma) \propto (\sigma^2)^{-(a+\frac{n}{2}+1)}\exp\left\{ - \frac{1}{2\sigma^2}\left(2b + \lVert\mbf{Y} - \mbf{X}\bgamma\be\rVert_2^2\right)\right\} \pi^{M-M_1} (1-\pi)^{M_1}f_\pi(\pi)
\end{equation}
where $\lVert\cdot\rVert_2$ is the $\ell_2$-norm. The E-step uses the current value of the parameters, denoted by $\bTheta^{(t)}$ in step $t$, to obtain the expected log-posterior 
\begin{equation}\label{eq.Qfunc}
Q(\bTheta|\bTheta^{(t)}) = E_{\bgamma}\left\{\log p(\bTheta|\mathcal{D},\bgamma)|\D, \bTheta^{(t)} \right\},
\end{equation}
which will be a function of $\mbf{p}=(p_1,\ldots,p_M)$ where $p_m=\Pr(\gamma_m=1|\D,\bTheta)$. The M-step consists of maximizing (\ref{eq.Qfunc}) to obtain $\bTheta^{(t+1)}$. The procedure is iterated until some convergence criteria is reached. 

Many Bayesian variable selection methods with the EM include $\gamma_m$ in the prior variance for $\beta_m$, where if $\gamma_m=1$, the prior variance is ``large'' and the prior variance is ``small'' otherwise. Our model formulation can be seen as the limit of this idea where the prior variance is $0$ or infinite depending on the $\gamma$ value. The benefit of our approach is that the values for ``large'' and ``small'' variances and the form of the prior distribution do not need to be specified. Some challenges with our model formulation can occur due to (i) Bayes factors favoring the null model for non-informative improper priors, and (ii) the posterior probability of $\gamma_m$ given $\beta_m$ is not independent of $\D$. When $\gamma_m$ is in the prior variance, the posterior distribution of $\gamma_m$ given $\beta_m$ is independent of $\D$. This simplifies derivation of the posterior expectation of $\gamma_m$,  but assumes it is independent of the dependence structure of the data. Our method relaxes this assumption through the use of parameter expansion.

Our goal is to develop an algorithm to obtain MAP estimates of  $\beta_m$ for $m=1,\ldots, M$ and $\sigma^2$ in the $M \gg n$ setting. Further, we provide an estimate of $\mbf{p}$, which is used with the MAP of $\be$ to estimate $\bgamma\be$. In Section \ref{sec.CM.step} and \ref{sec.m.step}, we show how the M-step can be completed given the expectations from the E-step. In Section \ref{sec.e.step}, we discuss how the E-step will be performed in light of the challenges discussed above.

\subsection{CM-step}\label{sec.CM.step}

In this Section, we demonstrate how to maximize the Q-function using the ECM algorithm. To simplify the presentation, we assume $\sigma^2$ and $\pi$ are known. Estimating these quantities is addressed later. The theory for the ECM was proposed in a seminal paper by \cite{MenRub93} who showed that the ECM is a generalized EM algorithm. The ECM algorithm can be used to simplify the optimization by replacing a single computationally complex M-step with $S$ computationally simpler CM-steps. Let $\bTheta$ be put into $S$ (possibly overlapping) sets $(\theta_1,\theta_2,\ldots,\theta_S)$ and $\bTheta^{(t+s/S)}$ denote the value of $\bTheta$ at the $s$th CM-step of iteration $t$. The value $\bTheta^{(t+s/S)}$ is chosen to maximize (\ref{eq.Qfunc}) under the constraint that $\bTheta$ is fixed at $\bTheta^{(t+(s-1)/S)}$ for all $\bTheta \notin \theta_s$. 

An important consideration is the choice of the sets such that it leads to computationally simple updates. Here, we set $\theta_s \equiv \theta_m = \beta_m$ for all $m=1,2,\ldots,M$ and assume the columns of $\X$ have been arranged in some updating order (discussed further in Section \ref{sec.estimation}). The first CM-step maximizes (\ref{eq.Qfunc}) respect to $\beta_1$ while holding $\be_{/1}=\be_{/1}^{(t)}$ which yields  
\begin{equation}\label{eq.simple.beta}
    \beta_1^{(t+1)} =(\X_1'\X_1)^{-1} \X_1'(\Y - \W_1^{(t)})
\end{equation}
where $\W_1^{(t)} = E_{\gamma_{/1}}(\X_{/1}\bgamma_{/1} \be_{/1}^{(t)}|\D, \bTheta^{(t)} )$ is an $n\times 1$ vector which will be a function of $\mbf{p}_{/1}$. The remaining CM-steps proceed in a similar manner; the $m$th CM-step will maximize  (\ref{eq.Qfunc}) respect to $\beta_m$ with the constraint that $\be_{/m}= (\beta_1^{(t+1)},\ldots, \beta_{m-1}^{(t+1)}, \beta_{m+1}^{(t)}, \ldots, \beta_M^{(t)})$.  The expectations are conditional on $\bTheta^{(t)}$ since the E-step commences after all of the CM steps. The multicycle ECM -- which interjects the CM-steps with updates of the E-step -- is not considered here. Note that the updates in (\ref{eq.simple.beta}) have the same form as those derived in variational Bayes with the CAVI formulation under Gaussian priors with infinite variance \citep[see][]{CarSte12}.

\subsection{PX-CM-step}\label{sec.m.step}

The Parameter-Expanded EM (PX-EM) algorithm is a variant that aims to improve the stability and convergence rates over the standard EM. \cite{Liuetal98} gave conditions where the PX-EM has the same convergence properties as the EM and ECM. The PX-EM expands the parameters under consideration to $\Gamma = (\bTheta_\ast, \alpha)$ where $\bTheta_\ast$ has the same dimension as $\bTheta$, $\bTheta = R(\bTheta_\ast,\alpha)$ for a known function $R$ where $\bTheta = R(\bTheta_\ast,\alpha_0)$ implies $\bTheta_\ast = \bTheta$ for some $\alpha_0$, and $\D$ contributes no information on the expanded parameter $\alpha$. Let $\Gamma^{(t)} = (\bTheta^{(t)}_\ast, \alpha^{(t)})$ and  $\bTheta^{(t)}= R(\bTheta^{(t)}_\ast,\alpha^{(t)})$ be the expanded and reduced parameters at iteration $t$ with $\Gamma^{(t)}_0 = (\bTheta^{(t)}, \alpha_0)$. The Q-function for the expanded parameter model is 
\begin{equation}\label{eq.Qfunc.px}
Q(\Gamma|\Gamma^{(t)}_0) = E_{\bgamma}\left\{p(\bTheta,\alpha|\mathcal{D},\bgamma)|\D, \Gamma^{(t)}_0 \right\}
\end{equation}
where $p(\bTheta,\alpha|\mathcal{D},\bgamma)$ is equal to (\ref{eq.Likelihood}) with $\alpha$ and $p(\alpha)$ included and $p(\bTheta|\mathcal{D},\bgamma) = p(\bTheta,\alpha_0|\mathcal{D},\bgamma)$. Throughout, we use $p(\alpha)\propto 1$ as the prior for $\alpha$. The PX-E-step calculates the expectation of (\ref{eq.Qfunc.px}). The PX-M-step contains two parts: find $\Gamma^{(t+1)}$ by maximizing (\ref{eq.Qfunc.px}), and mapping $\Gamma^{(t+1)}$ back to the original parameter space yielding $\bTheta^{(t+1)}= R(\bTheta^{(t+1)}_\ast,\alpha^{(t+1)})$.


We now discuss combining the PX-EM and ECM algorithms. Similar to Section \ref{sec.CM.step}, this consists of coordinate-wise maximization with the $\alpha$ parameter is included.  The $\alpha$ term is used to adjust for the impact of all predictors that have yet to be updated. To this end, we consider the following parameter expanded form of (\ref{eq.model})
\begin{equation}\label{eq.model.cmpx}
\Y =  \W_{m-} + \X_m \gamma_m\beta_m  +  \alpha \W_{m+} + \mbf{\epsilon}
\end{equation}
where $\W_{m-} = \sum_{k=1}^{m-1} \X_k \gamma_k\beta_k$, $\W_{m+} = \sum_{k=m+1}^{M} \X_k \gamma_k\beta_k$, and $\alpha_0=1$. For step $m<M$ in iteration $t$, the Q-function corresponding to (\ref{eq.model.cmpx}) is maximized with respect to $(\beta_m,\alpha)$ under the constraint that $\beta_k=\beta_{k}^{(t+(m-1)/M)}$ for all $k\neq m$ and $\sigma^2 = \sigma^{2(t)}$.  After maximization, we apply the reduction function to $\{\beta_k;k>m\}$, which gives $\beta_{k}^{(t+(m-1)/M)}$ (see (\ref{eq.remap})). Once a parameter has been optimized, it is fixed at that value for the remainder of the iteration. Note that for step $M$, $\W_{m+} = \mbf{0}$ and no $\alpha$ term is used. It is straightforward to show that the Q-function is non-decreasing in $m$.

More formally, let $\xi_1,\ldots,\xi_M$ denote the sets of the expanded parameter space where $\xi_m=(\beta_m,\alpha)$ for $m=1,\ldots,M-1$, and $\xi_{M}=(\beta_M,\sigma^2)$. For the $m$th PX-CM-step ($m<M$) maximizing the Q-function with respect to $(\beta_m,\alpha)$ gives $\bfeta^{(t+m/M)}_m \equiv (\beta_m^{{(t+m/M)}} \ \alpha^{(t+m/M)})' $
\begin{equation}\label{eq.beta.alpha}
\bfeta^{(t+m/M)}_m  = \left(\begin{array}{cc}
    \X_m' \X_m  & \X_m' \W_{m+}^{(t)} \\
     p_m^{(t)}\W_{m+}^{(t)'}\X_m & \mbf{1}'\W_{m+}^{2(t)}
\end{array} \right)^{-1} \left(\begin{array}{c}
     \X_m \\
      \W_{m+}^{(t)}
\end{array} \right)^\prime (\Y - \W_{m-}^{(t)})
\end{equation}
where $\W_{m+}^{(t)} = E_{\gamma_{/m}}(\W_{m+}|\D, \Gamma_0^{(t+(m-1)/M)})$, and $\W_{m+}^{2(t)} = E_{\gamma_{/m}}\{(\W_{m+})^2|\D, \Gamma_0^{(t+(m-1)/M)}\}$.   To obtain (\ref{eq.beta.alpha}) we assumed $p_m^{(t)}>0$. When $p_m^{(t)}=0$, there is no information to inform the value of $\beta_m$. However, we can still calculate $\bfeta^{(t+m/M)}_m$ by using $\lim_{p_m^{(t)}\rightarrow 0}\bfeta^{(t+m/M)}_m$.  Reverse mapping results in new values of $\W_{m+1-}^{(t)}$, $\W_{m+1+}^{(t)}$, and $\W_{m+1+}^{2(t)}$ which are used in the following PX-CM-step (see Section \ref{sec.mom.W}). This process is completed for $m=1,\ldots,M-1$. The $M$th step has no $\alpha$ term and results in $\beta_M^{(t+1)} = \X_M'(\Y - \W_{M-}^{(t)})/(\X_M'\X_M)$, which is then used to update $\sigma^2$ with 
\begin{eqnarray}\label{eq.sigma2.update}
 \sigma^{2(t+1)} = \frac{2b^{(t)}_p}{n-1},
\end{eqnarray}
where $b^{(t)}_p = \left\{\Y'\Y -2\Y'(W_{M-}^{(t)}+p_M\beta_M^{(t+1)}\X_m) + \mbf{1}'W_{M-}^{2(t)} +p_M\X_m'\X_m\beta_M^{(t+1)2}\right\}/2$.  Section A of the Supplemental Material contains derivations of the above PX-CM step updates.

\subsection{E-step}\label{sec.e.step}

As discussed in the introduction, a main issue with using flat uninformative priors is that Bayes factors comparing null and alternative models will favor the null model, regardless of the data. That is, the Bayes factor comparing a model with $\gamma_m=0$ to one with $\gamma_m=1$ (with $\bgamma_{/m}$ at the true value) will equal zero with the proposed prior favoring the null model. As a result, Bayes factors cannot be used to complete the E-step. Below we show how methods motivated from the multiple testing literature can be used to approximate the expectations of $\gamma_m$'s. We then discuss the forms and computations of the expectations required to compute the PX-CM steps.

\subsubsection{Two-groups approach}\label{sec.two.groups}

Our alternative to Bayes factors for estimating $p_m$ uses an estimate of the posterior variance of $\beta_m$ at step $t$ for all $m$. For now, assume $S_m^{(t)}$ and $S_m^{2(t)}$ are estimates of the posterior standard deviation and variance of $\beta_m|\gamma_m=1$, respectively, which we discuss in Section \ref{sec.post.var}. Consider the distribution of $\beta_m$ conditional on $(\mathcal{D},\W_{m-},\W_{m+})$, which we refer to as the \textit{oracle} posterior. It is straightforward to show that the oracle posterior is Gaussian with mean zero if $\gamma_m=0$ and non-zero mean if $\gamma_m=1$. The goal is to use the approximated posterior distribution at step $t$, i.e., $\beta_m \sim N(\beta_m^{(t)},S_m^{2(t)})$, to determine if the oracle posterior distribution has zero or non-zero mean. That is, does the approximated posterior arise from an oracle posterior that has a mean of zero? Note the similarities of this problem with the popular `two-group' approach to multiple testing \citep[e.g.,][]{Efron2001,Sun2007,Sto07,Efr08} where statistics (e.g., means and variances) are available for a large number of tests which have a mixture of zero and non-zero expectations. Estimation of the posterior probability that the null is true (i.e., mean is zero) conditional on a test statistic that is equal to an observed value (the so-called ``local FDR'') has received considerable attention within the two-group approach.  Empirical Bayes approaches have been shown to have desirable properties  \citep{Ste17,CasRoq20}. The empirical Bayes estimates are based on test statistics ${T}^{(t)}_m=\beta^{(t)}_m/S^{(t)}_m$, which is beneficial since we do not need to assume equal posterior variances, nor that the data have been decorrelated. 

Following the two-groups approach, we assume ${T}^{(t)}_m\sim (1-\gamma_m)f_0(\cdot) + \gamma_m f_1(\cdot)$ where $f_0(\cdot)$ is a standard normal distribution and $f_1$ is unknown since it is a function of $f_\beta$. This motivates the empirical Bayes estimate of the posterior expectation for $\gamma_m$,
\begin{equation}\label{eq.delta}
p^{(t)}_m = 1-\frac{\hat \pi^{(t)}_0 f_0({T}_m^{(t)})}{\hat{f}^{(t)}({T}_m^{(t)})},
\end{equation}
where $\hat{\pi}^{(t)}$ is the estimated proportion of null hypotheses, and $\hat{f}^{(t)}$ is the estimated marginal distribution of ${\mbf{T}}^{(t)}= ({T}^{(t)}_1,\ldots,{T}^{(t)}_M)$. To estimate $\pi$ we use the approach of \cite{Storey2004}, where $\hat{\pi}^{(t)} = \sum_m I({P}^{(t)}_m\geq \lambda)/\{M(1-\lambda)\}$ and ${P}_m^{(t)} = 2\{1-\Phi(|{T}^{(t)}_m|)\}$. For the tuning parameter, we use $\lambda=0.1$ as it has been demonstrated to have better properties with dependent data \citep{Blanchard2009}.  For $\hat f$ we use a kernel density estimate as $f^{(t)}(T)=(Mh)^{-1}\sum_m \phi\left(\frac{T-T_m^{(t)}}{h}\right)$ and assume (\ref{eq.delta}) is non-negative and monotone non-increasing from zero. For the bandwidth, all settings of the simulation use five times  Silverman's `rule of thumb' \citep{Sil86}, but for larger $M$ with weaker signals, larger bandwidths can be beneficial.  Note that $p^{(t)}_m = 0$ is a possibility with this procedure. We note that these are common choices for estimating $\pi$ and $f$, and many other approaches are available \citep[][among others]{JinCai07,RoqVil11,OstNic12}.

\subsubsection{Moments of $W$ and remapping}\label{sec.mom.W}

To obtain the moments required to complete the PX-CM-steps we first consider estimating the first and second moment of $\W_{\cdot} \equiv \sum_{k=1}^{M} \X_k \gamma_k\beta_k$. Using the exchangability of the $\gamma$'s, $\mbf{p}^{(t)}$, and $\be^{(t)}$, the first moment is given by $\W_{\cdot}^{(t)} = \X (\mbf{p}^{(t)} \be^{(t)} )$. The second moment is $\W^{2(t)}_{\cdot} = \mbf{V}_{\cdot}^{(t)} + (\W^{(t)}_{\cdot})^2$ where $\mbf{V}_{\cdot}^{(t)}$ is an $n \times 1$ vector with elements
\begin{eqnarray}\label{eq.Xt2}
{V}_{i\cdot}^{(t)} = Var(\X_{i}\bgamma \be^{(t)}|\D, \bTheta^{(t)} ) = \X_{i}^2 \left\{\be^{(t)2}  \mbf{p}^{(t)}(1- \mbf{p}^{(t)}) \right\}, \ \ i=1,\ldots,n. 
\end{eqnarray}
For $m=1$, the moments are given by $\W_{1-}=\mbf{0}$, $\W^{(t)}_{1+}  = \W_{\cdot}^{(t)} - \X_{1} {p}_{1}^{(t)} \beta_{1}^{(t)} $, $\W^{2(t)}_{1-} = \mbf{0}$, and $\W^{2(t)}_{1+} = \mbf{V}_{1+}^{(t)} + (\W^{(t)}_{1+})^2$ where $\mbf{V}_{1+}^{(t)} = \mbf{V}_{\cdot}^{(t)} - \X_{1}^2 \beta_{1}^{(t)2} {p}_{1}^{(t)}(1- {p}_{1}^{(t)})$. Let $\W^{2(t)}_{m+} = \mbf{V}_{m+}^{(t)} + (\W^{(t)}_{m+})^2$, similarly for $\W^{2(t)}_{m-}$ for all $m$.

The moments for step $m+1$ are calculated after step $m$ has been completed and the parameters have been remapped. After $\bfeta^{(t+m/M)}_m$ is obtained, reverse mapping yields 
\begin{eqnarray}\label{eq.remap} 
\beta_k^{(t+m/M)} = \left\{\begin{array}{ll}
    \beta_k^{(t+(m-1)/M)} & \mbox{ for } k<m  \\
    \alpha^{(t+m/M)}\beta_k^{(t+(m-1)/M)} & \mbox{ for } k>m
\end{array} \right. .
\end{eqnarray}
For step $m+1$, the moments are be updated via $\W_{m+1-}=\W_{m-} + \X_{m} p_{m}^{(t)}\beta_{m}^{(t+m/M)}$, $\W^{(t)}_{m+1+}  =  \alpha^{(t+m/M)}\W^{(t)}_{m+} - \X_{m+1} p_{m+1}^{(t)}\beta_{m+1}^{(t+m/M)}$, $\mbf{V}_{{m+1}-}^{(t)} = \mbf{V}_{m-}^{(t)} + \X_{m}^2 \beta_{m}^{(t+m/M)2} {p}_{m}^{(t)}(1- {p}_{m}^{(t)}) $, and $\mbf{V}_{{m+1}+}^{(t)} = \alpha^{(t+m/M)2}\mbf{V}_{m+}^{(t)} - \X_{m+1}^2 \beta_{m+1}^{(t+m/M)2} {p}_{m}^{(t)}(1- {p}_{m}^{(t)})$. The online remapping is computationally straightforward, and the high-dimensional matrix multiplication only needs to be completed once per iteration.


\subsection{Estimating the posterior variance}\label{sec.post.var}

The plug-in empirical Bayes estimators of hyperparameters require estimates of the posterior variance, denoted by $S_m^{2}$ for predictor $m$. There have been many approaches proposed to estimate the variability with the EM \citep{Lou82,MenRub91,Jametal00}, most of which involve numerical differentiation to estimate the observed information matrix. Here, we estimate the marginal posterior variance using the standard conditional variance relationship
\begin{equation}\label{eq.cond.var.true}
    S_m^{2} = Var(\beta_m|\D) = E\left\{Var(\beta_m|\W_{m+},\W_{m-})|\D\right\} + Var\left\{E(\beta_m|\W_{m+},\W_{m-})|\D\right\}
\end{equation}
\citep{Geletal14}. Using standard results, the complete data posterior is $(\beta_m,\alpha)|\W_m \sim N\{(\Z_m'\Z_m)^{-1}\Z'(\Y - \W_{m-}),\sigma^2 (\Z_m'\Z_m)^{-1} \}$ where $\Z_m = (\X_m \ \W_{m+})$. As a result, (\ref{eq.cond.var.true}) is the sum of the expectation of $(\Z_m'\Z_m)^{-1}$ and the variance of $\beta_m^{2(t+m/M)}$ (over $\W_{m+}$ and $\W_{m-}$), neither which have amendable closed form. Consequently, we estimate (\ref{eq.cond.var.true}) using the first-order Taylor series approximation which yields
\begin{equation}\label{eq.cond.var}
    S_m^{2} \approx S_m^{2(t+1)} = Var(\beta_m|\W^{(t)}_{m-}, \W^{2(t)}_{m-}) + (\mbf{b}_{m+}^{(t)})^{2\prime}\mbf{V}_{m+}^{(t)}+ (\mbf{b}_{m-}^{(t)})^{2\prime}\mbf{V}_{m-}^{(t)},
\end{equation}
where $\mbf{b}_{m+}^{(t)} = \partial E(\beta_m|\W_{m-},\W_{m+})/\partial\W_{m+}$ evaluated at $\W_{m-} = \W_{m-}^{(t)}$ (similarly for $\mbf{b}_{m-}^{(t)}$).  In Section C of the Supplemental Material, we show that $Var(\beta_m|\W^{(t)}_{m-}, \W^{2(t)}_{m+})$ can be estimated using the first diagonal element of
\begin{equation}\label{eq.first.var}
    \sigma^{2(t)} \left(\begin{array}{cc}
    \X_m' \X_m  & \X_m' \W_{m+}^{(t)} \\
     \W_{m+}^{(t)'}\X_m & \mbf{1}'\W_{m+}^{2(t)}
\end{array} \right)^{-1},
\end{equation}
and the $i$th row of $\mbf{b}_{m+}^{(t)}$ is
$$ b_{im+}^{(t)} = H^{-1}\left\{
    2W_{im+}^{(t)}(\Y_{m}^{(t)\prime}\X_m)-Y^{(t)}_{im}(\W_{m+}^{(t)\prime}\X_m) -X_{im}(\Y_{m}^{(t)\prime}\W_{m+}^{(t)}) - h_i \beta_m^{(t+m/M)}\right\},$$
where $\Y_{m}^{(t)} = \Y - \W_{m-}^{(t)}$, $h_i = 2(W_{mi}^{(t)}\X_m'\X_m - p_mX_{im}\X_m'\W_m^{(t)}) $ and $H = (\X_m'\X_m)$ $(\W_m^{(t)\prime} \W_m^{(t)}) - p_m(\X_m'\W_m^{(t)})^2$. Similarly, $b_{im-}^{(t)} = H^{-1}\left\{X_{im} \mbf{1}'\W_{m+}^{2(t)} - W_{im+}^{(t)}(\W_{m+}^{(t)\prime}\X_m)\right\}$. For all $m$, (\ref{eq.cond.var}) is used to estimate the posterior variances which are then used to complete the empirical Bayes portion of the E-step.

\subsection{All-at-once version}\label{sec.all.at.once}

\begin{figure}[t]
\centering
\begin{tabular}{ccc}
 \includegraphics[width=5in]{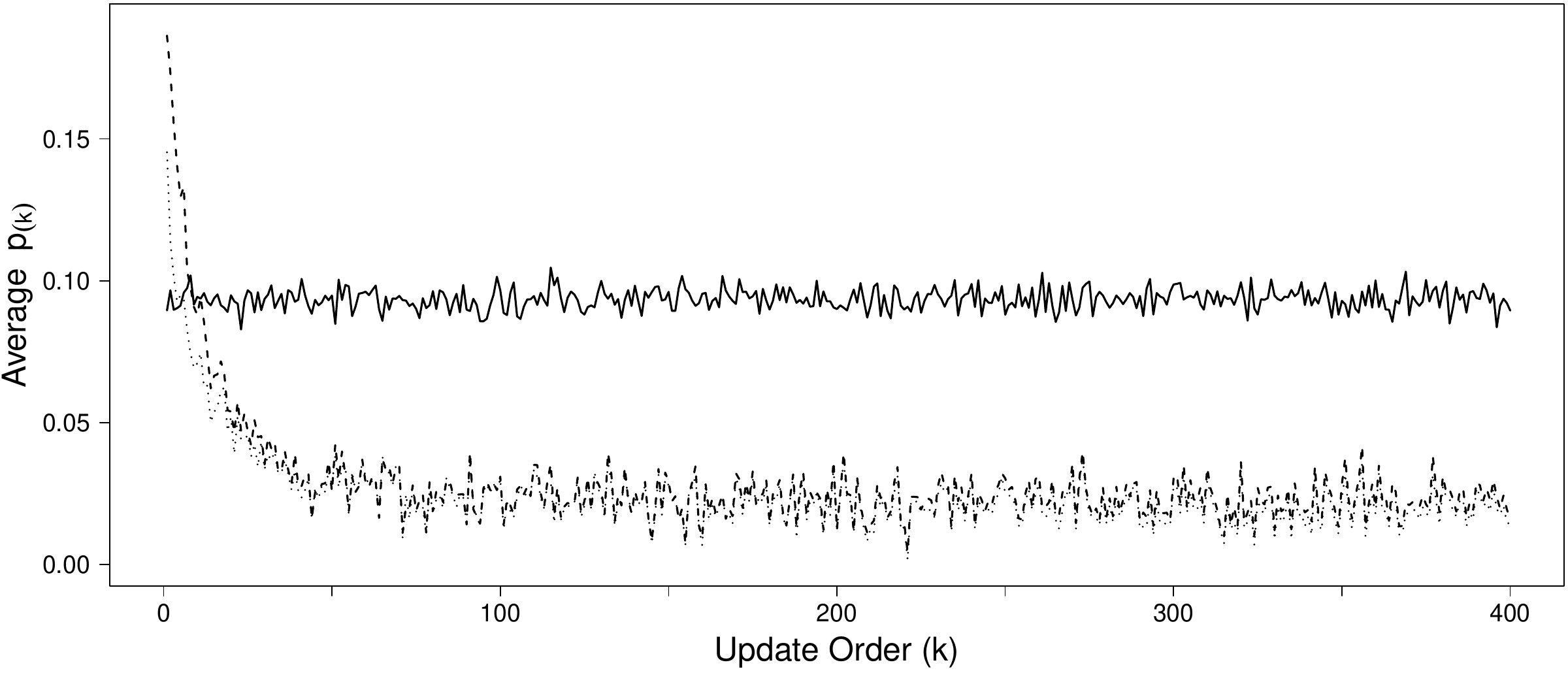} 
\end{tabular}
\vspace{-0.5cm}
\caption{Average $p_{(k)}$ by updating order for one-at-a-time PROBE (dotted line), sparse variational Bayes with Laplace priors (dashed line), and all-at-once PROBE (solid line). Updating order is chosen at random; $p_{(1)}$ is updated first, then $p_{(2)}$, etc..  \label{fig.p.by.k}}
\end{figure}

The methods outlined above give the basis for the one-at-a-time PROBE algorithm (given fully in Section \ref{sec.estimation}. This algorithm can be sensitive to the order of the updates, particularly for dependent data, where the predictors with low update order (i.e., those that are updated early) will have artificially inflated $\beta$ and $p$. This occurs because predictors with low update order can absorb much of the impact of all related predictors that are updated thereafter \citep{CarSte12}. In Figure \ref{fig.p.by.k}, we demonstrate this phenomenon with simulated data where the updating order is chosen at random (note: this choice of updating order is only for demonstration purposes). The figure shows the average $p_{(k)}$ where $p_{(1)}$ is updated first, $p_{(2)}$ second, etc., for one-at-a-time PROBE, a sparse variational Bayes with Laplace priors (another one-at-a-time procedure), and the all-at-once PROBE (described below). The predictors updated earlier have the first chance to explain the unexplained variability in the model. This causes the coefficients updated first to have larger $p_{(k)}$ with corresponding $\beta_{(k)}$ that tend to have positive absolute biases (results not shown). On the other hand, the all-at-once PROBE gives all predictors the same opportunity to account for any unexplained variation in the outcome by updating all predictors as though they were first in the updating order.

We now outline the all-at-once variant of the PROBE algorithm. The CM-steps for the all-at-once version proceed similarly to the one-at-a-time version, with each predictor being updated as $m=1$ is updated. This implies that $\W_{m-}=\mbf{0}$  and ${\W}_{m+} \equiv {\W}_m = \sum_{k \neq m} \X_k \gamma_k\beta_k$ for all $m$, with $\W_m^{(t)}$ and $\W_m^{2(t)}$ defined as appropriate. These are used to estimate $\beta_m^{(t+1)}$ via (\ref{eq.beta.alpha}) for all $m$, with the convention of setting $p^{(t)}_m=1$ when evaluating (\ref{eq.beta.alpha}). The reason for this convention is that the increase in variance due to the uncertainty of the E-step will be driven by the impact previous values of the parameters have on the E-step \citep{Oak99}. That is, the increase in posterior variance of $\beta_m^{(t+1)}$ is driven by $\partial \mbf{p}^{(t)}/\partial \beta_m^{(t)}$. Note that $\beta_m^{(t)}$ has a relatively large impact on $p^{(t)}_m$ and minimal impact on $p^{(t)}_k$ for $k\neq m$ (which would only be through the estimates of $\hat{\pi}^{(t)}$ and $\hat{f}^{(t)}$). Consequently, using the convention that $p^{(t)}_m=1$ minimizes the increase in posterior variance due to the E-step. For this reason, in the all-at-once algorithm $S_m^{2(t+1)}$ is set to the first diagonal element of (\ref{eq.first.var}), as the impact of the uncertainty of the E-step has been minimized. 

The update of $\sigma^2$ uses (\ref{eq.sigma2.update}) with $b^{(t)}_p = (\mathbf{Y}^{\prime}\mathbf{Y}- 2\alpha^{(t+1)}\mathbf{Y}^{\prime}\W^{(t)}_{\cdot} + \alpha^{(t+1)2}\mbf{1}'\W^{2(t)}_{\cdot})/2$ where
\begin{eqnarray*}
\alpha^{(t+1)} = \frac{\W_{\cdot}^{(t)'}\Y}{\mbf{1}'\W^{2(t)}_{\cdot}}.
\end{eqnarray*}
The estimation of $\mbf{p}^{(t)}$ is unchanged.   The updates of the moments of $\W$ proceed similarly where $\mbf{W}^{(t)}_{\cdot}$ and $\mbf{W}^{2(t)}_{\cdot}$ are calculated in the E-step and lead to computationally straightforward online calculations of $\mbf{W}^{(t)}_{m}$ and $\mbf{W}^{2(t)}_{m}$ for all $m$.

\section{Estimation}\label{sec.estimation}

We measure convergence in the algorithm via the change in consecutive $\W^{(t)}_{\cdot}$. Small changes in consecutive $\W^{(t)}_{\cdot}$ indicate that all estimated quantities are converging while relegating predictors with $p_{m}^{(t)}\approx 0$. We estimate convergence at iteration $t$ via $CC^{(t)} = \log(n)\max_i\{(W^{(t)}_{i \cdot} - W_{i \cdot}^{(t-1)})^2/V_{i \cdot}^{(t-1)}\}$ where $\log(n)$ offsets the impact of sample size on the maximum of Chi-squared random variables \citep{Embetal13}. In our simulations and data analyses, we use $CC^{(t)} < \chi^2_{1,\varepsilon}$, where $\chi^2_{1,\varepsilon}$ represents the $\varepsilon$ quantile of a Chi-squared distribution with $1$ degree of freedom, as the stopping criterion. We use $\varepsilon= 10^{-3}$ and $10^{-1}$ for the all-at-once and one-at-a-time variants, respectively. 

\begin{algorithm}[t]
\caption{One-at-a-time PROBE algorithm.}\label{algo.1}
\begin{algorithmic}[]
 \State  Initiate $\mbf{W}^{(0)}$ and $\mbf{W}^{2(0)}$
    \While{$CC^{(t)} \geq \chi^2_{1,\varepsilon} $ and $\max(\mbf{p}^{(t)})>0$} 
   \State  \textbf{PX-CM-step}
   \For{$m = 1,2,\ldots,M-1$}
        \State \textbf{Maximization:} use $\mbf{W}_m^{(t)}$, and $\mbf{W}_m^{2(t)}$ to estimate $\bfeta^{(t+m/M)}_m$ via (\ref{eq.beta.alpha}).
        \State \textbf{Reduction:} re-map parameters according to (\ref{eq.remap}).
        \State Online updating of $\mbf{W}_{m+1}^{(t)}$ and $\mbf{W}_{m+1}^{2(t)}$.
        \State Calculate $S_m^{(2t+1)}$ via (\ref{eq.cond.var}).
        \EndFor
        \State \hspace{2mm} Use $\mbf{W}_M^{(t)}$, and $\mbf{W}_M^{2(t)}$ to estimate $\beta^{(t+1)}_M$, $S_M^{2(t+1)}$, and $\sigma^{2(t+1)}$ using (\ref{eq.sigma2.update}).
        \State Limit step size of $\beta^{(t+1)}_m$ and $S^{2(t+1)}_m$ using (\ref{eq.mixing}) for all $m$.
\State \textbf{PX-E-step}
    \State \hspace{2mm} (a) Estimate $\hat{f}^{(t+1)}$ and $\hat{\pi}^{(t+1)}_{0}$ and use them to calculate $\mbf{p}^{(t)}$ via (\ref{eq.delta}).
    \State \hspace{2mm} (b) Calculate $\mbf{W}^{(t+1)}_\cdot$ and $\mbf{W}^{2(t+1)}_\cdot$.
\State Calculate $CC^{(t+1)}$ and check convergence.
    \EndWhile  
\end{algorithmic}
\end{algorithm}

The steps of one-at-a-time PROBE are given in Algorithm \ref{algo.1}. To choose the order of the PX-CM steps, Lasso regression is performed on the standardized data, and the order is set according to the absolute value of the estimated coefficients. A similar approach was used by \cite{RaySza22}. To start the algorithm, we set $\be^{(0)}$ to the estimated coefficients from a least absolute shrinkage and selection operator \citep[LASSO,][]{Tib96} model with the penalty parameter that minimized the prediction error via $10$-fold cross-validation (CV), $\mbf{p}^{(0)}=\mbf{1}$, and $\sigma^{2(0)}=s^2_y$ the sample variance of $\Y$.

The full version of all-at-once PROBE is given in Algorithm \ref{algo.A}. By design, all-at-once PROBE is not sensitive to the updating order since all predictors are updated as if they were the first. To start the all-at-once algorithm, we set $\be^{(0)} = \mbf{0}$, $\mbf{S}^{(0)} = \mbf{0}$, $\mbf{p}^{(0)}=\mbf{0}$, and $\sigma^{2(0)}=s^2_y$ the sample variance of $\Y$. For this approach, ${{\beta}}^{(1)}_m$ is the simple linear regression coefficient of $\X_m$ on $\Y$ for all $m$. In the simulations and data analyses, if the algorithm results in $\mbf{p}^{(t)}=\mbf{0}$ for some $t$ we restart with the alternative approach, where if $\mbf{p}^{(t)}=\mbf{0}$ again the algorithm stops and null values are returned for the MAP estimates. 


\subsection{Limiting step size}

For both versions of the algorithm, we limit the step sizes from one iteration to the next.  Specifically, let $\hat \beta^{(t+1)}_m$ and $\hat S^{2(t+1)}_m$ denoted the estimated values of $\beta_m$ and $ S^2_m$ at iteration $t+1$. These values and the estimates from the previous iteration are used to update $\beta_m$ and $ S^2_m$ via
\begin{eqnarray}\label{eq.mixing}
\beta^{(t+1)}_m &=& (1-q^{(t+1)})\beta^{(t)}_m  + q^{(t+1)}{\hat \beta}^{(t+1)}_m, \ \mbox{and} \nonumber \\ 
S^{2(t+1)}_m &=& \{(1-q^{(t+1)})(S^{2(t)}_m)^{-1} + q^{(t+1)}(\hat{S}_m^{2(t+1)})^{-1}\}^{-1},
\end{eqnarray}
respectively. \cite{PetWal78} and \cite{TitWan06} studied the convergence properties of similar updates for normal mixture models \citep[see also,][]{Deletal99}. We apply (\ref{eq.mixing}) between the M- and E-steps. \cite{VarRol08} and \cite{Jiaetal22} proposed similar modifications to accelerate the convergence of the EM. 

The $q^{(t)}$ acts as a sequence of learning rates or step sizes -- $q^{(t)}=1$ will ignore the previous estimates, and $q^{(t)}=0$ will ignore current estimates. Large learning rates can result in estimates that repeatedly oscillate or cycle between values, inhibiting convergence. This phenomenon has also been observed in Expectation Propagation \citep[EP,][]{Min01} where damping the EP updates has been proposed \citep{MinLaf02,Vehetal20}. Momentum, in gradient descent and other deep learning approaches, has a similar function \citep{Qia99,Sutetal13}. In the simulations and data analyses, we use $q^{(t)}=(t+1)^{-1}$ for the all-at-once method and $q^{(t)}=(t+1)^{-0.5}$ for the one-at-a-time method, which we found to have good properties in all settings.

\begin{algorithm}[t]
\caption{All-at-once PROBE algorithm.}\label{algo.A}
\begin{algorithmic}[]
 \State  Initiate $\mbf{W}^{(0)}$ and $\mbf{W}^{2(0)}$
    \While{$CC^{(t)} \geq \chi^2_{1,\varepsilon} $ and $\max(\mbf{p}^{(t)})>0$} 
   \State  \textbf{M-step}
        \State \hspace{0.6 cm} Use $\mbf{W}^{(t)}_m$ and $\mbf{W}^{2(t)}_m$ to estimate $\beta^{(t+1)}_m$ and ${{S}}_m^{2(t+1)}$ for $m=1,\ldots,M$.
        \State \hspace{0.6 cm} Use $\mbf{W}^{(t)}_{\cdot}$ and $\mbf{W}^{2(t)}_{\cdot}$ to estimate $\alpha^{(t+1)}$ and $\sigma^{2(t+1)}$.
        \State  Limit step size of $\beta^{(t+1)}_m$ and $S^{2(t+1)}_m$ using (\ref{eq.mixing}) for all $m$.
\State \textbf{E-step}
    \State\hspace{0.6 cm} (a) Estimate $\hat{f}^{(t+1)}$ and $\hat{\pi}^{(t+1)}_{0}$ and use them to calculate $\mbf{p}^{(t+1)}$ via (\ref{eq.delta}).
    \State\hspace{0.6 cm} (b) Calculate $\mbf{W}^{(t+1)}_{\cdot}$ and $\mbf{W}^{2(t+1)}_{\cdot}$.
\State Calculate $CC^{(t+1)}$ and check convergence.
    \EndWhile  
\end{algorithmic}
\end{algorithm}

At convergence we obtain the MAP estimates $(\tilde{\be}, \tilde{\mbf{p}}, \tilde{\sigma}^2)$ where $\tilde \sigma^2 = 2\tilde{b}_p/(n-1)$, $\tilde b_p = (\mathbf{Y}^{\prime}\mathbf{Y}- 2\mathbf{Y}^{\prime}\tilde{\W}_{\cdot} + \mbf{1}'\tilde{\W}^{2}_{\cdot})$, and $(\tilde{\W}_{\cdot},\tilde{\W}^{2}_{\cdot})$ are obtained from the final iteration. In our simulation studies and data analyses, we use $\tilde{p}_m \tilde \beta_m $  to estimate $\gamma_m\beta_m$ for all $m$.

\subsection{Theoretical Considerations and Computational Complexity}\label{sec.theo}

The one-at-a-time PROBE algorithm is a quasi-PX-ECM algorithm. What differentiates it from a standard PX-ECM algorithm is the completion of the E-step.  Plug-in empirical Bayes values of the hyperparameters were used in the E-step because of \textit{Barlett's paradox}, i.e., Bayes factors always favoring the null with the chosen prior. As a result, a standard PX-ECM is not possible. A benefit of the empirical Bayes procedure is that $M$ will be large, and the data-driven estimates of $\pi$ and $f$ will be based on sufficient sample size. The all-at-once PROBE algorithm alters the one-at-a-time version to overcome its dependence and sensitivity on the updating order. These changes further remove it from a standard PX-ECM algorithm (and its theoretical guarantees). In general, the convergence properties of all-at-once type optimization procedures are more difficult to obtain than their one-at-a-time counterparts. However, as demonstrated below, the performance of all-at-once PROBE is superior to the one-at-a-time version.  

The computational complexity of the all-at-a-once PROBE algorithm is such that it can incorporate a large number of predictors. Each cycle of the EM requires $M+1$ linear regression models, where equation (\ref{eq.beta.alpha}) has complexity $O(q^2N+q^3)$ where $q=2$ when there is no intercept and $q=3$ with an intercept (as we use in our simulation studies and data analyses). The updates of the E-step require $O(NM)$ complexity (this can be reduced when $p_m=0$ for some $m$). In total, the computational complexity for $K$ EM-steps is $O\{K(M+1)(N+q^2N+q^3)\}$. In comparison, $K$ iterations of the LASSO has $O\{KN(M+k)\}$ where $k$ is the number of non-zero coefficients \citep{Hasetal15}. As a result, the computational complexity grows linearly with $N$ and $M$ for both methods. Below, we compare the computation times of both PROBE algorithms with various alternative approaches for several data scenarios.

\section{Simulation Studies}\label{sec.sim}

To empirically test the performance of the PROBE algorithm, we performed numerous simulation studies. The outcome was generated via 
$Y_i = \X_i\bgamma\be  +   \epsilon_{i} $
where $\be \sim \mathcal{U}(0,2\eta)$, and $\epsilon_i \sim N(0,\sigma^2)$. We used continuous and binary $\X$ on a two dimensional $\sqrt{M} \times \sqrt{M}$ grid where $\mbf{d}_m = (d_{1m},d_{2m}) \in (1,\ldots,\sqrt{M})^2$ denotes the coordinates of predictor $m$. For the continuous setting $\X_{i} \sim^{iid} MVN(0,\Sigma_{s})$ where  $\Sigma_{s}(\mbf{d}_m,\mbf{d}_{m^\prime}) = \exp\{-\lVert (\mbf{d}_m - \mbf{d}_{m^\prime})/s \rVert_2^2\}$, and $s=10$. The binary setting used $\X^b_{i} = I\{ \X_{i}< 0\}$ where $I(\cdot)$ denotes the indicator function. For $\bgamma$, we set $\gamma_m = I\{G_m<Q_{\pi}\}$ where $\mbf{G} \sim MVN(0,\Sigma_{20})$ and $Q_{\pi}$ was such that $M_1/M=\pi$. 

This data generation scheme results in dependence within $\X$ and $\bgamma$. For $\X$, $\Sigma_{s}$ has decaying covariance away from the (unit) diagonal. Thus, the dependence between predictors ranges from strong to (essentially) zero, similar to real data. Additionally, because $\bgamma$ is randomly generated in each iteration, the results average over a range of possible true non-null configurations, which include one, two, or more clusters of non-null predictors (see Section C of the Supplementary Materials for examples of $\bgamma\be$ and $\X$). The settings were varied over all combinations of $M\in (20^2,50^2,10^4)$, $\pi \in (0.01,0.05,0.1)$, $\eta \in (0.3,0.5,0.8)$, and $\sigma^2$ was such that the signal-to-noise ratio $SNR = Var(\X_i\bgamma\be)/\sigma^2 \approx 1$ or $2$. All simulations used $n=400$ data points and were ran for $B=1000$ iterations.

Algorithms \ref{algo.1} and \ref{algo.A} were used to fit one-at-a-time and all-at-once versions of PROBE. The comparison methods included LASSO, adaptive LASSO \citep[ALASSO,][]{Zou06}, minimax concave penalty \citep[MCP,][]{Zha10}, Ridge regression \citep{HoeKen70}, smoothly clipped absolute deviation penalty \citep[SCAD,][]{FanLi01}, along with the EM for spike-and-slab \citep[EMVS,][]{RocGeo14}, spike-and-slab LASSO \citep[SSLASSO,][]{RocGeo18}, sparse variational Bayes \citep[SPARSEVB,][]{RaySza22}, VB for spike-and-slab priors \citep[VARBVS,][]{CarSte12}, and an MCMC empirical Bayes approach using Gaussian slab priors \citep[EBREG,][]{Maretal17}. To fit the penalization methods, the penalty parameter that minimized the $10$-fold cross-validated (CV) MSE was found using \texttt{glmnet} \citep[for LASSO and ALASSO,][]{Frietal10} and \texttt{ncvreg} \citep[for SCAD and MCP,][]{BreHua11}. To fit the Bayesian methods, the settings of the hyperparameters used values relatively consistent with the default settings for the packages, and (at times) the true values of some parameters were used (see Section A of the Supplementary Material for full details). In practice, these parameters could be further refined to improve performance. The EBREG results are based on $100$ iterations due to their computational complexity, and it was not included with $M=400$ due to issues with setting the prior when LASSO results in a null model. The results from Ridge regression and EMVS are not shown as they were not competitive with the other methods.

\begin{figure}[t]
\centering
\begin{tabular}{ccc}
 \includegraphics[scale=0.57]{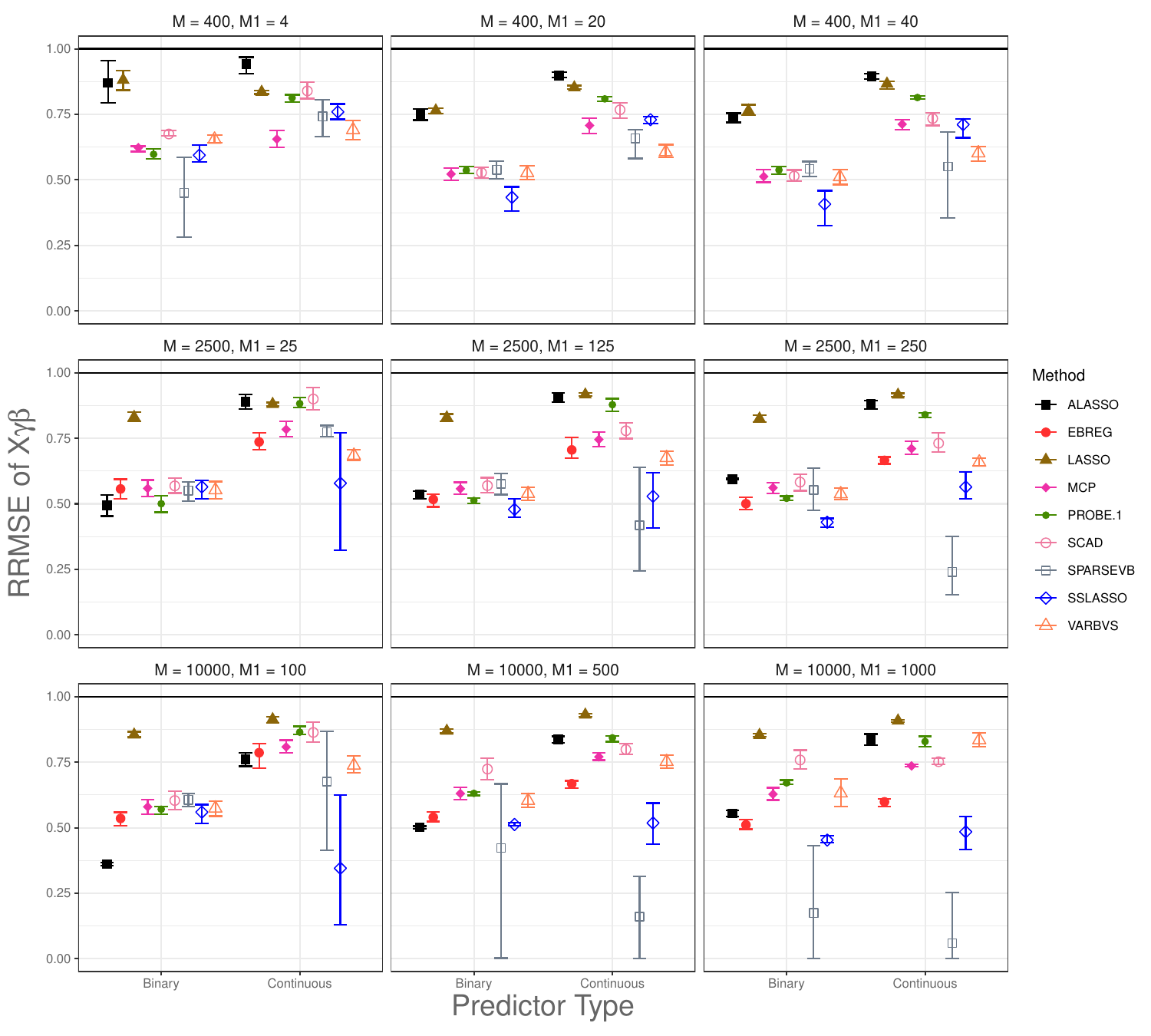} 
\end{tabular}
\vspace{-0.2cm}
\caption{RMSE of $\X\bgamma\be$ for all-at-once PROBE relative the RMSE of comparable methods ($RRMSE = RMSE_{\textsc{probe}_\textsc{a}}/RMSE_{\textsc{method}}$) by $M$, $M_1$, and the predictor type. Bars represent the range (min/max) and points the average $RRMSE$ over all $\sigma^2$ and $\eta$ combinations.\label{fig.mse.sig}}
\end{figure}

\begin{figure}[t]
\centering
\begin{tabular}{c}
 \includegraphics[scale=0.57]{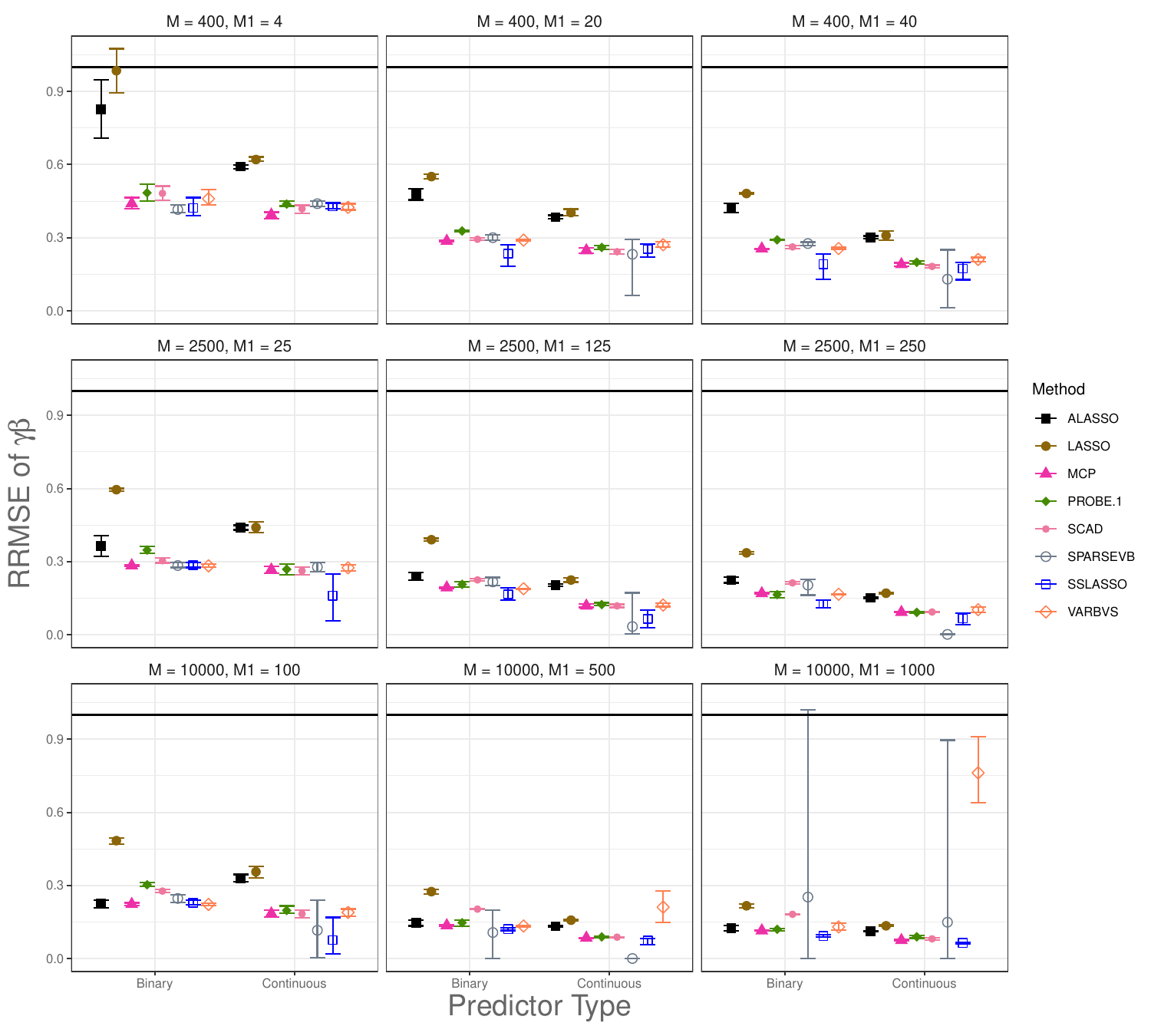} 
\end{tabular}
\vspace{-0.2cm}
\caption{RMSE of $\gamma_m\beta_m$ for all-at-once PROBE relative the RMSE of comparable methods ($RRMSE = RMSE_{\textsc{probe}_\textsc{a}}/RMSE_{\textsc{method}}$) by $M$, $M_1$, and the predictor type. Bars represent the range (min/max) and points the average $RRMSE$ over all $\sigma^2$ and $\eta$ combinations.\label{fig.mse.beta}}
\end{figure}

The performance of the models was quantified with the root of the average mean squared error (RMSE) of estimates of $\X_i\bgamma\be$ (over all $i$) and $\gamma_m\beta_m$ (over all $m$). To facilitate comparisons over the many situations that were considered, we utilize the relative RMSE (RRMSE). For setting $s$ and method $j$,  RRMSE is defined by
$$RRMSE_{\textsc{method}_j}^s = \frac{{RMSE_{\textsc{probe}_\textsc{a}}^s}}{{RMSE_{\textsc{method}_j}^s}},$$
where $RMSE_{\textsc{probe}_\textsc{a}}^s$ is the RMSE (of $\X_i\bgamma\be$ or $\gamma_m\beta_m$) for setting $s$ over all $B$ iterations for all-at-once PROBE, and $RMSE_{\textsc{method}_j}^s$ is the corresponding value for method $j$.  Having $RRMSE_{\textsc{method}_j}^s<1$ indicates all-at-once PROBE was more efficient than method $j$ for setting $s$ (all-at-once PROBE is not included in the figures since its relative MSE is one by definition). 

\subsection{Simulation Results}

Figures \ref{fig.mse.sig} and \ref{fig.mse.beta} display the range and average relative RMSE (RRMSE) for $\X_i\bgamma\be$ and $\gamma_m\beta_m$, respectively, over all settings for the given $M$, $M_1$ and predictor type.  The RMSEs of all-at-once PROBE for $\X_i\bgamma\be$ were lower than all comparable methods for each setting tested. The closest comparison was with ALASSO for $M=400$, $M_1=4$, $\eta=0.5$, $SNR=2$ and continuous predictors ($RRMSE=0.97$).  Overall, LASSO had the second-lowest RMSE. The estimates of $\gamma_m\beta_m$ from all-at-once PROBE had the lowest RMSE in 104 of the 108 simulation settings (see Figure \ref{fig.mse.beta}). The LASSO had a lower RMSE in three settings with $M=400$ and $M_1=4$, and SPARSEVB had the lowest in one setting with $M=10^4$ and $M_1=10^3$. For most settings, the $\gamma_m\beta_m$ RMSE for all-at-once PROBE is three to four times lower than the RMSEs of the comparable methods. The results for one-at-a-time PROBE are similar to those from SCAD, MCP, and EBREG. The performance of all-at-once PROBE was superior to the one-at-a-time variant in any metric for all settings.

\begin{figure}[t]
\centering
\begin{tabular}{ccc} 
  \includegraphics[scale=0.6]{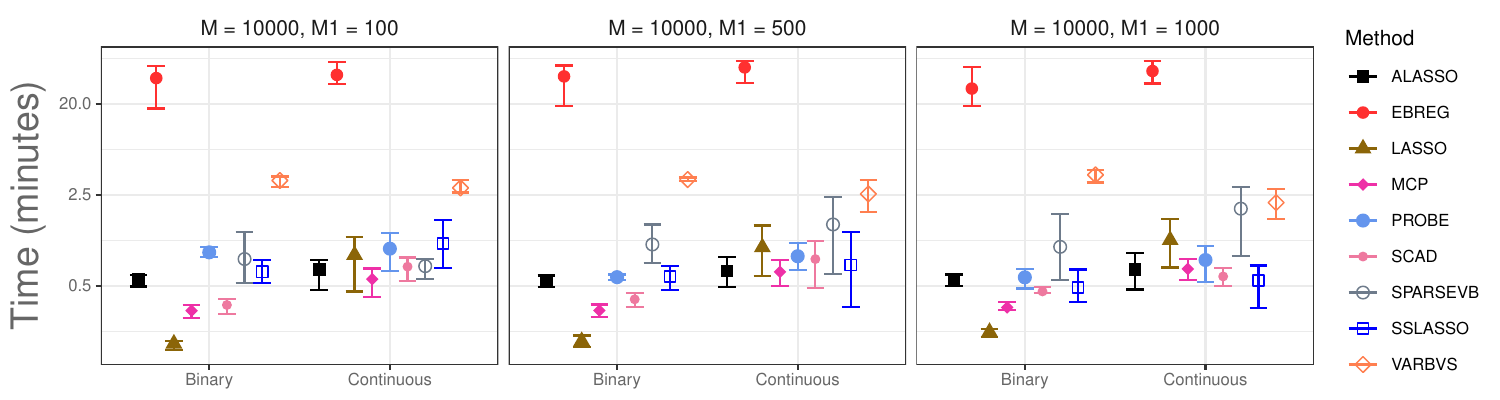} 
\end{tabular}
\vspace{-0.5cm}
\caption{Average estimation time for $M=10^4$ (on $\log$ scale). Bars represent the range (min/max) and points the average over all variations of the given setting.\label{fig.sigma}}
\end{figure}

Figure \ref{fig.sigma} shows the range and mean average estimation time (in minutes) for $M=10^4$. For these settings, LASSO, MCP, and SCAD had the quickest estimation time.  The estimation time for all-at-once PROBE was similar to that of SPARSEVB and SSLASSO, both of which are considered to be computationally efficient Bayesian methods. Section A of the Supplementary Materials contains figures comparing estimation times for other $M$ values and the relative median absolute deviation. Code to reproduce the simulations can be found at \cite{McLZgo21}.

\section{Analysis cancer cell lines on drug response}\label{sec.anal}

We demonstrate both the one-at-a-time and all-at-once PROBE variants on a study analyzing drug response in cancer cell lines. The data was obtained from the Cancer Cell Line Encyclopedia \citep[CCLE,][]{Barretina2012}. Currently, this database contains 136,488 datasets covering a total of 1457 cell lines. In particular, data measuring responses to 24 anticancer drugs were available. After removing missing data, 18,988 predictors were available (all consisted of human gene expression levels) from eight drug response outcomes. We limited our analysis to these eight drugs and defined the area above the dose-response curve as the drug response outcome, following \cite{Dondelinger2018}. Section B of the Supplementary Materials provides more details, and files to reproduce all results can be found at \cite{McLZgo21}. 

\begin{figure}[t]
\centering
\begin{tabular}{c}
\includegraphics[width=6in]{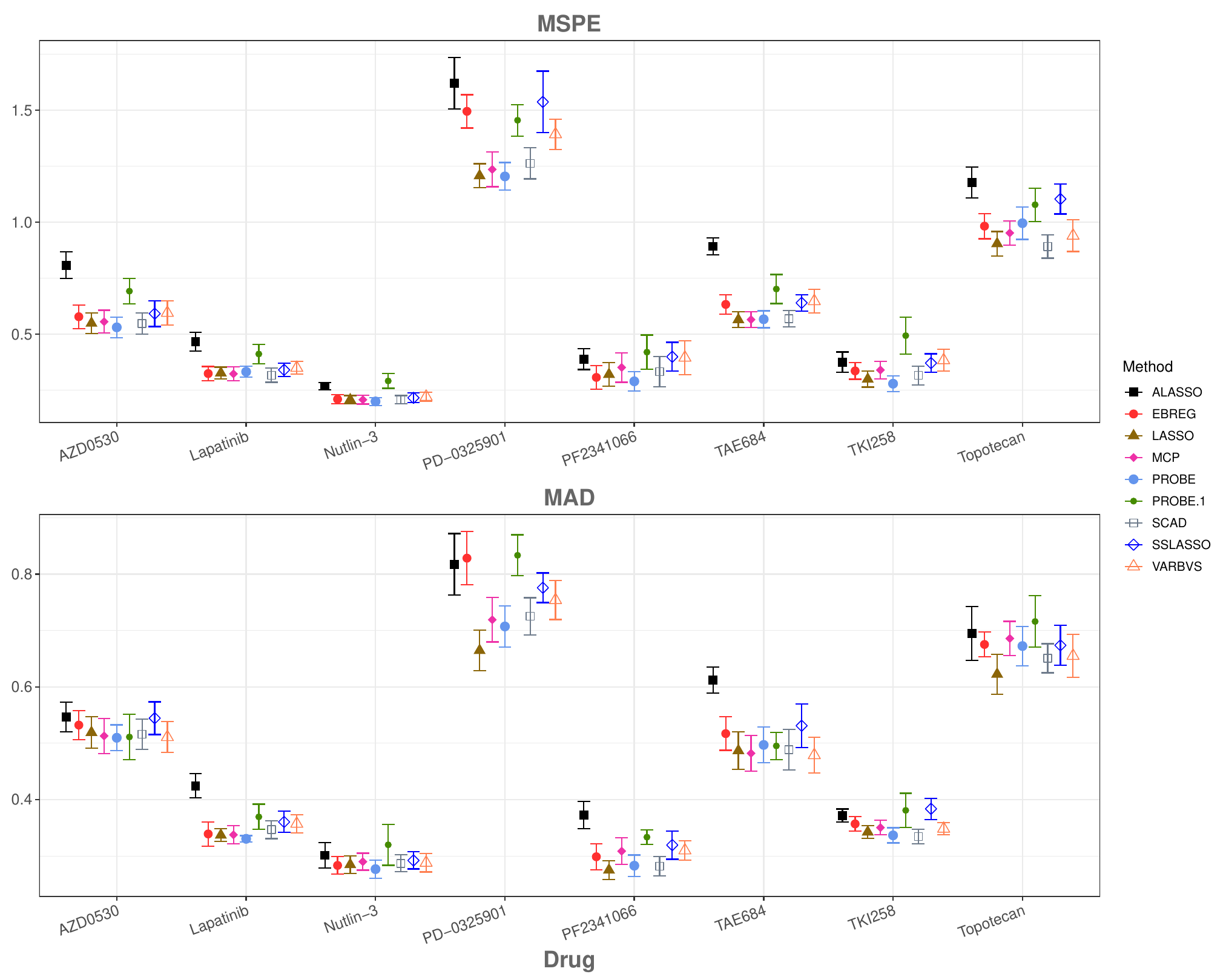}
\end{tabular}
\caption{(top) CV-MSPE and (bottom) CV-MAD for eight anticancer drug response models for the six predictive methods. Displayed is the overall MSPE $\pm$ the estimated standard error (all based on 10-fold CV). \label{fig.mse.example1}}
\end{figure}

All methods used in the simulation section were applied to the data. For the Bayesian methods, the hyperparameters were selected using default approaches. Predictive ability was quantified with CV mean squared predictive error (MSPE) and median absolute deviation (MAD). Both PROBE methods used the same stopping criterion and learning rate as in Section \ref{sec.sim}. To simplify the presentation, we do not display prediction errors from SPARSEVB (similar figures with SPARSEVB can be viewed in the Supplementary Material).

The MSPE and MAD results are shown in Figure \ref{fig.mse.example1}. All-at-once PROBE had the lowest MSPE for five drugs (AZD0530, Nutlin-3, PD-0325901, PF2341066, TKI258), while SCAD (Lapatinib, Topotecan) and LASSO (TAE684) had the lowest for the remaining drugs. Regarding MAD, all-at-once PROBE and LASSO had the lowest for three drugs each (AZD0530, Nutlin-3, Lapatinib, and PD-0325901, PF2341066, Topotecan, respectively), while SCAD (TKI258) and VARBVS (TAE684) had the lowest for the remaining drugs.

\section{Discussion}\label{sec.disc}

The proposed PROBE method is a novel estimation strategy for sparse high-dimensional linear models. The Bayesian sparse linear model framework is minimally influenced by prior model assumptions. The key aspects of the proposed estimation procedure are the use of the partitioned ECM framework with expanded parameters and plug-in empirical Bayes estimates of hyperparameters. Two variants of the methodology were proposed. The one-at-a-time variant is closer to a PX-ECM algorithm but differentiated by the use of an empirical Bayes E-step. This E-step was used because Bayes factors will favor the null model when non-informative priors are used. The main motivation for the all-at-once version is to remove the impact of the updating order on the resulting estimates. For the one-at-a-time type of optimization methods, the updating order must be supplied, and the results can be sensitive to this step. We demonstrated that the estimates of $p_m$ can be larger for variables updated first, even when chosen at random. 

In the simulation studies, all-at-once PROBE performed better than the one-at-a-time version, penalization methods, and other Bayesian variable selection methods. While PROBE requires running $M+1$ linear models at each iteration of the algorithm, the program is not overly computationally intensive. The statistical code utilizes \texttt{RcppArmadillo} \citep{EddSan14}, which results in rapid CM-step calculations. In the simulation studies with $M=10^4$, all-at-once PROBE had similar computation times to popular penalization approaches. One-at-a-time PROBE required more steps to converge than all-at-once and was one of the slower procedures tested.

In this study, we have applied the PROBE algorithm to high-dimensional sparse linear regression. The PROBE framework could be expanded to robust regression, mixed models, or generalized linear models in the high-dimensional setting. Further, the algorithm could be altered such that blocks of predictors are simultaneously updated in the M-step, which would be useful for predictors with natural groupings (i.e., interactions or nominal predictors).  Work is currently underway to explore such possibilities. An \texttt{R} \citep{R} package \texttt{probe}, code to replicate simulation studies and the data analyses, and examples with further adjustment variables are available on Github \citep{McLZgo21}.


\bibliographystyle{asa}
\bibliography{PROBE}

\newpage

\begin{center}
{\LARGE \bf Supplementary Material} 
\end{center}

\section*{A. Derivation of PX-CM step updates}\label{app.pxcm.steps}

In this Appendix, we derive the form for the updates of $\bfeta^{(t+m/M)}_m$ for $m<M$ along with $\beta_M^{(t+1)}$ and $\sigma^{2(t+1)}$ given in the main text. Recall the parameter-expanded form of the model
\begin{equation}\label{eq.model.cmpx.app}
\Y =  \W_{m-} + \X_m \gamma_m\beta_m  +  \alpha \W_{m+} + \mbf{\epsilon}
\end{equation}
where $\W_{m-} = \sum_{k=1}^{m-1} \X_k \gamma_k\beta_k$, $\W_{m+} = \sum_{k=m+1}^{M} \X_k \gamma_k\beta_k$, and $\alpha \sim p(\alpha)\propto 1$. The log-posterior of this model is given by 
\begin{eqnarray}\label{eq.log.post}
\log{p(\Gamma|\mathcal{D},\bgamma)}  &=&  - \frac{1}{2\sigma^2}\left(2b + \lVert\mbf{Y} - (\W_{m-} + \X_m \gamma_m\beta_m  +  \alpha \W_{m+})\rVert_2^2\right)  \nonumber \\ 
&&  -\left(a+\frac{n}{2}+1\right)\log(\sigma^2)
+ M\log(\pi)+ M_1\log\left(\frac{1-\pi}{\pi}\right) \nonumber \\
&& + \log\{f_\pi(\pi)\} +C
\end{eqnarray}
where $\Gamma = (\be,\sigma^2,\pi, \alpha)$ and $C$ is used to denote a constant that is independent of the parameters of interest. 

The value $\bfeta^{(t+m/M)}_m$  for $m<M$ maximizes the Q-function corresponding to (\ref{eq.log.post}) with respect to $(\beta_m,\alpha)$ while holding $\beta_k=\beta_{k}^{(t+(m-1)/M)}$ for all $k\neq m$. This Q-function is given by 
\begin{eqnarray}\label{eq.q.func}
Q(\Gamma|\Gamma^{(t+(m-1)/M)})  &=& E\left( \lVert\mbf{Y} - (\W_{m-} + \X_m \gamma_m\beta_m  +  \alpha \W_{m+})\rVert_2^2\big|\mathcal{D}, \Gamma^{(t+(m-1)/M)}\right)  + C \nonumber \\
&=& - 2p_m\beta_m\mbf{Y}_m^{(t)\prime}\X_m - \alpha 2\mbf{Y}_m^{(t)\prime}\W_{m+}^{(t)} - 2p_m\beta_m\alpha\X_m'\W_{m+}^{(t)}  \nonumber \\ 
&& + p_m\beta_m^2\X_m' \X_m + \alpha^2 \mbf{1}'\W_{m+}^{2(t)},
\end{eqnarray}
where $\mbf{Y}_m^{(t)} = \mbf{Y} - \W^{(t)}_{m-}$. Taking the derivatives with respect to $(\beta_m,\alpha)$ gives $\Dot{\mbf{Q}}(\Gamma)_m = (\frac{\partial}{\partial \beta_m}, \frac{\partial}{\partial \alpha})'Q(\Gamma|\Gamma^{(t+(m-1)/M)})$ given by 
\begin{eqnarray}\label{eq.q.derv}
\Dot{\mbf{Q}}_m(\Gamma) = \left ( \begin{array}{l}
     -2p_m\mbf{Y}_m^{(t)\prime}\X_m  - 2p_m\alpha\X_m'\W_{m+}^{(t)} + 2p_m\beta_m\X_m' \X_m \\
      -  2\mbf{Y}_m^{(t)\prime}\W_{m+}^{(t)} - 2p_m\beta_m\X_m'\W_{m+}^{(t)} + 2\alpha \mbf{1}'\W_{m+}^{2(t)} 
\end{array} \right).  \nonumber
\end{eqnarray}
Setting $\Dot{\mbf{Q}}_m(\Gamma)=\mbf{0}$ and solving yields
\begin{equation}\label{eq.beta.alpha.app}
\left(\begin{array}{c}
     \beta_m  \\
     \alpha
\end{array}\right)  = \left(\begin{array}{cc}
    \X_m' \X_m  & \X_m' \W_{m+}^{(t)} \\
     p_m^{(t)}\W_{m+}^{(t)'}\X_m & \mbf{1}'\W_{m+}^{2(t)}
\end{array} \right)^{-1} \left(\begin{array}{c}
     \X_m \\
      \W_{m+}^{(t)}
\end{array} \right)^\prime (\Y - \W_{m-}^{(t)}), \nonumber
\end{equation}
as the maximum values.  

For $\beta_M$, there is no $\alpha$ term since $\W_{M+} = \mbf{0}$. The remaining Q-function for $(\beta_M,\sigma^2)$ is 
\begin{eqnarray}\label{eq.log.post.M}
Q(\Gamma|\Gamma^{(t+(M-1)/M)}) =  -\left(a+\frac{n}{2}+1\right)\log(\sigma^2) - \frac{1}{2\sigma^2}\left(2b + \lVert\mbf{Y} - \W_{M-} - \X_M \gamma_M\beta_M \rVert_2^2\right). \nonumber 
\end{eqnarray}
Using the same notation as above, solving $\Dot{\mbf{Q}}_M(\Gamma)=\mbf{0}$ for $\beta_M$ yields $\beta_M^{(t+1)} = \X_M'(\Y - \W_{M-}^{(t)})/(\X_M'\X_M)$ as the maximum value. The derivative with respect to $\sigma^2$ gives
\begin{eqnarray}\label{eq.der.sig}
\sigma^2 = \frac{b + \frac{1}{2}E(\lVert\mbf{Y} - \W_{M-} - \X_M \gamma_M\beta_M^{(t+1)} \rVert_2^2\big|\mathcal{D}, \Gamma^{(t+(M-1)/M)})}{a+\frac{n}{2}+1} \nonumber 
\end{eqnarray}
as the maximum value. Plugging in $a=-3/2$ and $b=0$ and evaluating the expectation yields
\begin{eqnarray}\label{eq.sigma2.update.app}
 \sigma^{2(t+1)} = \frac{2b^{(t)}_p}{n-1}, \nonumber
\end{eqnarray}
where $b^{(t)}_p = \left\{\Y'\Y -2\Y'(W_{M-}^{(t)}+p_m\beta_M^{(t+1)}\X_m) + \mbf{1}'W_{M-}^{2(t)} +p_m\X_m'\X_m\beta_M^{(t+1)2}\right\}/2$.

\section*{B. Derivation of posterior variances}\label{app.post.var}

In this Section, we derive the form for $S_m^{2(t+1)}$ given in the main text. Recall that, using the first-order Taylor series approximation $S_m^{2(t+1)}$ is estimated with
\begin{equation}\label{eq.cond.var.app}
    S_m^{2(t+m/M)} = Var(\beta_m|\W^{(t)}_{m-}, \W^{2(t)}_{m-}) + (\mbf{b}_{m+}^{(t)})^{2\prime}\mbf{V}_{m+}^{(t)}+ (\mbf{b}_{m-}^{(t)})^{2\prime}\mbf{V}_{m-}^{(t)},
\end{equation}
where $\mbf{b}_{m+}^{(t)} = \partial E(\beta_m|\W_{m-},\W_{m+})/\partial\W_{m+}$ evaluated at $\W_{m-} = \W_{m-}^{(t)}$ (similarly for $\mbf{b}_{m-}^{(t)}$).  

We first derive the form for $Var(\beta_m|\W^{(t)}_{m-}, \W^{2(t)}_{m-})$ which is the posterior variance of $\beta_m$ given $\W_{m-}$ and $\W_{m+}$ evaluated at $\W^{(t)}_{m-}$ and $ \W^{2(t)}_{m+}$. Letting $\Z_{m} = (\X_m, \W_{m+})$, the posterior distribution of $\beta_m$ given $\W_{m-}$ and $\W_{m+}$ is the mixture
\begin{eqnarray*}\label{eq.post.eta}
\beta_m|\mathcal{D},\sigma^2, \gamma_m \sim \gamma_m N(\hat{\beta}_m, S^2_{m1} ) + (1-\gamma_m)\delta_0(\cdot)
\end{eqnarray*}
where $(\hat{\beta}_m, \hat{\alpha}) = ({\Z}_m'{\Z}_m)^{-1}\Z_m'(\Y - \W_{m-})$, and $S^2_{m1}$ denotes the $(1,1)$ element of ${\sigma}^2({\Z}_m'{\Z}_m)^{-1}$. In the main text, we plug in $(\W^{(t)}_{m-}, \W^{2(t)}_{m-})$ and use $Var(\beta_m|\mathcal{D}) = S^2_{m1}$ for all $m$, thereby using the posterior variance of $\beta_m|\gamma_m=1$. This will tend to overestimate the variance compared to using the expected variance but avoids potential issues when $p_m=0$.  

To obtain the derivative of $\beta_m$, note that given $\W_{m-}$ and $\W_{m+}$, the value of $\beta_m$ that maximizes the Q-function is
\begin{equation}\label{eq.beta.der}
H^{-1} \{(\W_{m+}^{\prime} \W_{m+})\{\X_m'(\Y - \W_{m-})\} - (\X_m\W_{m+}') \{\W_{m+}'(\Y - \W_{m-})\}\nonumber
\end{equation}
where $H = (\X_m'\X_m) (\W_{m+}^{\prime} \W_{m+}) - p_m(\X_m'\W_{m+})^2$. Taking the derivative with respect to $\W_{m-}$ and $\W_{m+}$ and using the first order approximation yields the form in Section \ref{sec.post.var}.

\section*{C. Additional Simulation Details and Results}

The following settings were used to fit the Bayesian variables selection methods (the notation corresponds to the package documentation). To fit EMVS we used a sequence from $-2$ to $2$ with increments of $0.25$ for $\log(\nu_0)$, $\nu_1=1000$, a fixed prior on $\theta$ with $\theta=M_1/M$ (the true value), and a conjugate prior on $\be$ and $\sigma^2$ (results correspond to $\nu_0=e^{-2}$). For SSLASSO we used $\lambda_1=0.1$, $\lambda_0$ a series between $\lambda_1$ and $M_1$ with 400 elements, $a=M_1$ and $b=M-M_1$ with unknown variance. For VARBVS the residual variance and prior variance of the regression coefficients were approximated using maximum-likelihood. For SPARSEVB, Laplace and Gaussian prior distributions were fit (results for Laplace were marginally better and were used), the true value of $\sigma$ was used for the residual standard deviation, and an intercept was included. For EBREG we used $\alpha=0.99$, $\gamma=0.005$, a prior for $\sigma^2$ and $\log\{I(|\bgamma| \leq n)/n\}$ as the log-prior on the model size.

\begin{figure}[t]
\centering
\begin{tabular}{ccc}
 \includegraphics[scale=0.6]{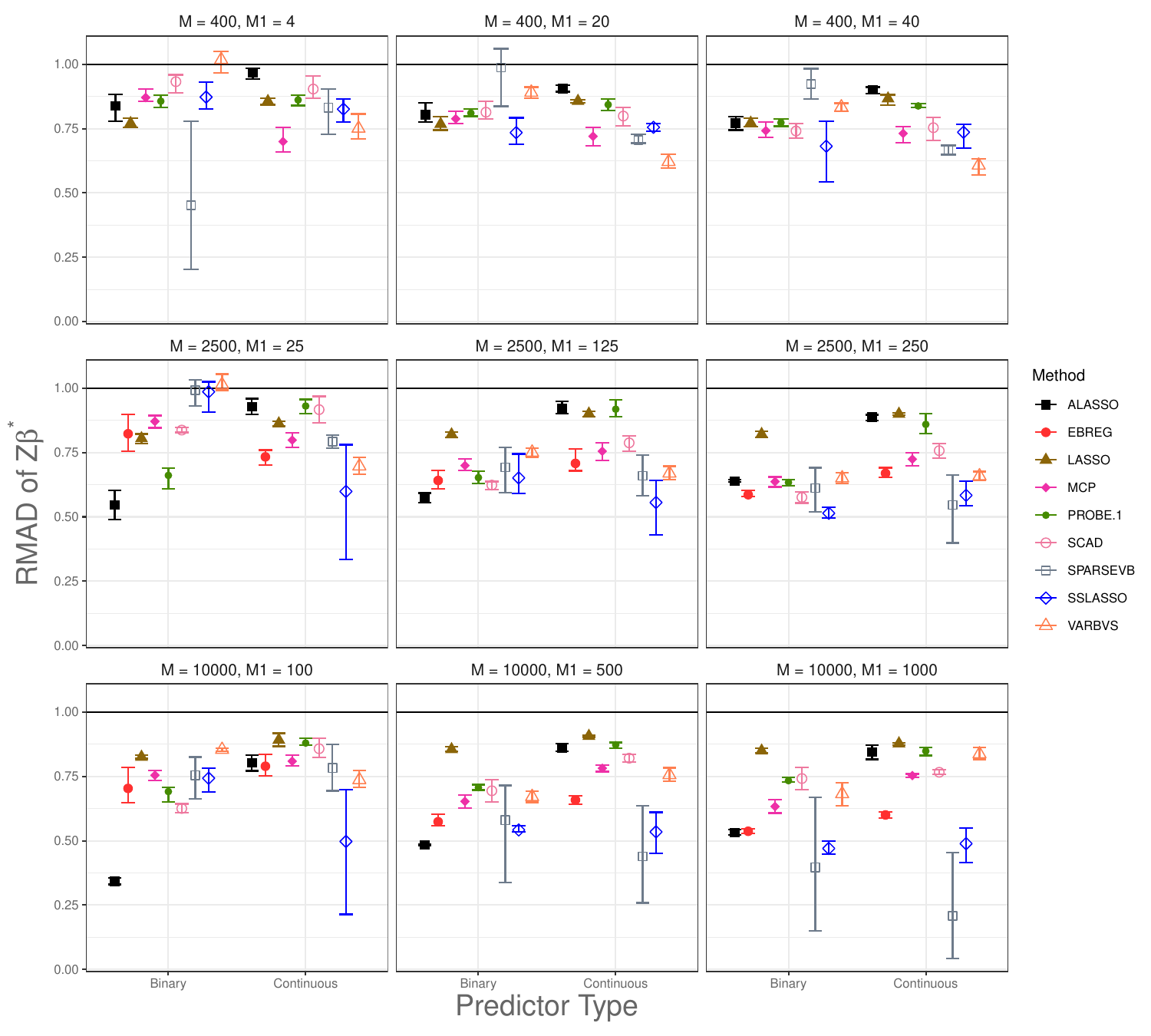} 
\end{tabular}
\vspace{-0.2cm}
\caption{MAD of $\X\bgamma\be$ for PROBE relative the MAD of competing methods (MAD$_{\textsc{probe}_\textsc{a}}$/MAD$_{\textsc{method}}$) by $M$, $M_1$, and the predictor type. Bars represent the range (min/max) and points the average relative MAD over all SNR and $\eta$ settings. \label{fig.MAD}}
\end{figure}

\begin{figure}[t]
\centering
\begin{tabular}{ccc}
 \includegraphics[scale=0.6]{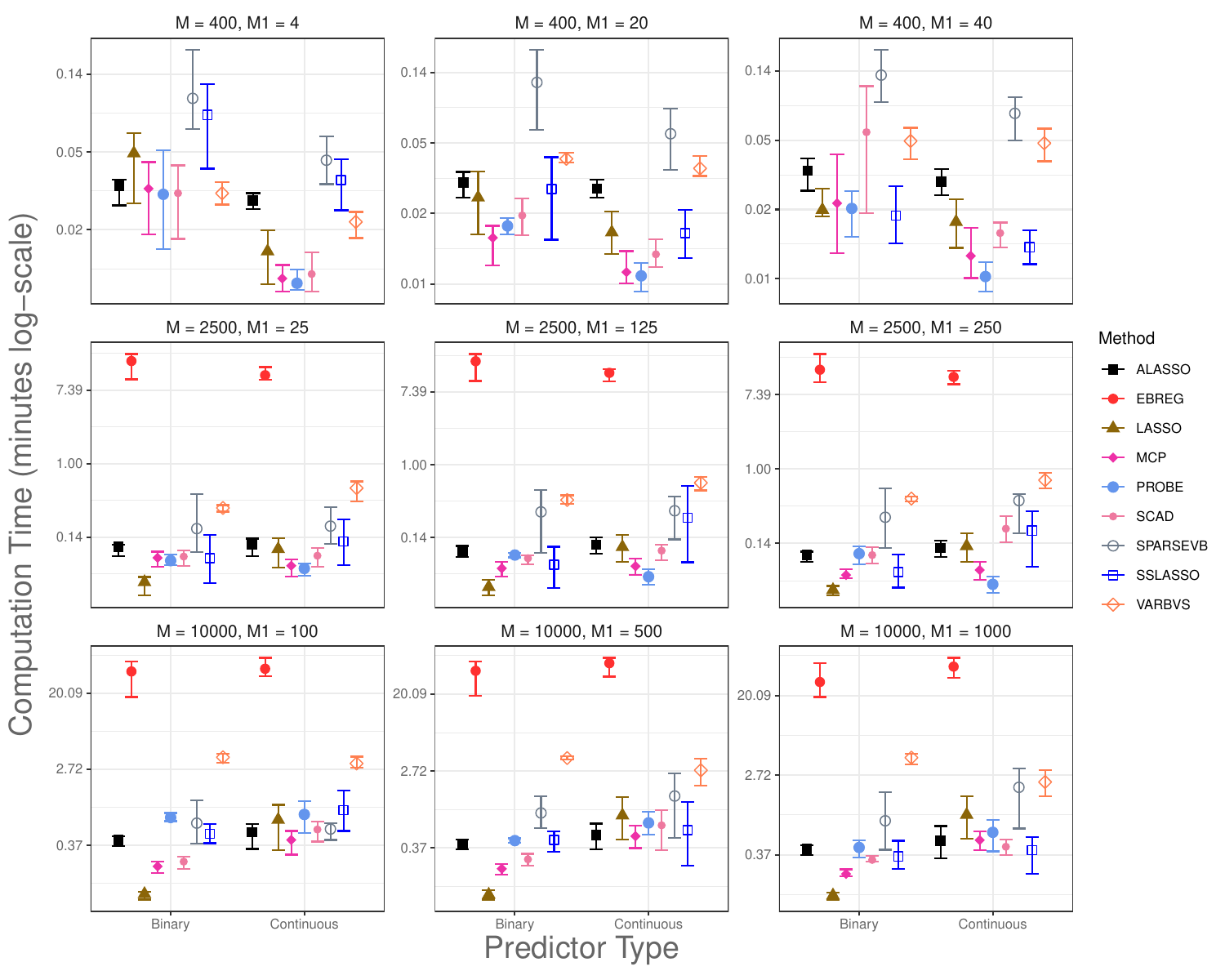} 
\end{tabular}
\vspace{-0.2cm}
\caption{Average time (in seconds) transformed by $\log_{10}$. Bars represent the range (min/max) and points the average over all $SNR$ and $\eta$ combinations.\label{fig.time.res}}
\end{figure}

Figure \ref{fig.MAD} contains results on the median absolute deviation (MAD) of predictions of $\X\bgamma\be$. Displayed is the MAD of PROBE divided by the MAD of the comparison method (i.e., values $< 1$ indicate PROBE was more efficient). Figure \ref{fig.time.res} shows the range and mean average estimation time (in seconds) transformed by $\log_{10}$ for all $M$ (main text only showed this for $M=10^4$).

\begin{figure}[t]
\centering
\begin{tabular}{ccc}
 \includegraphics[scale=0.35]{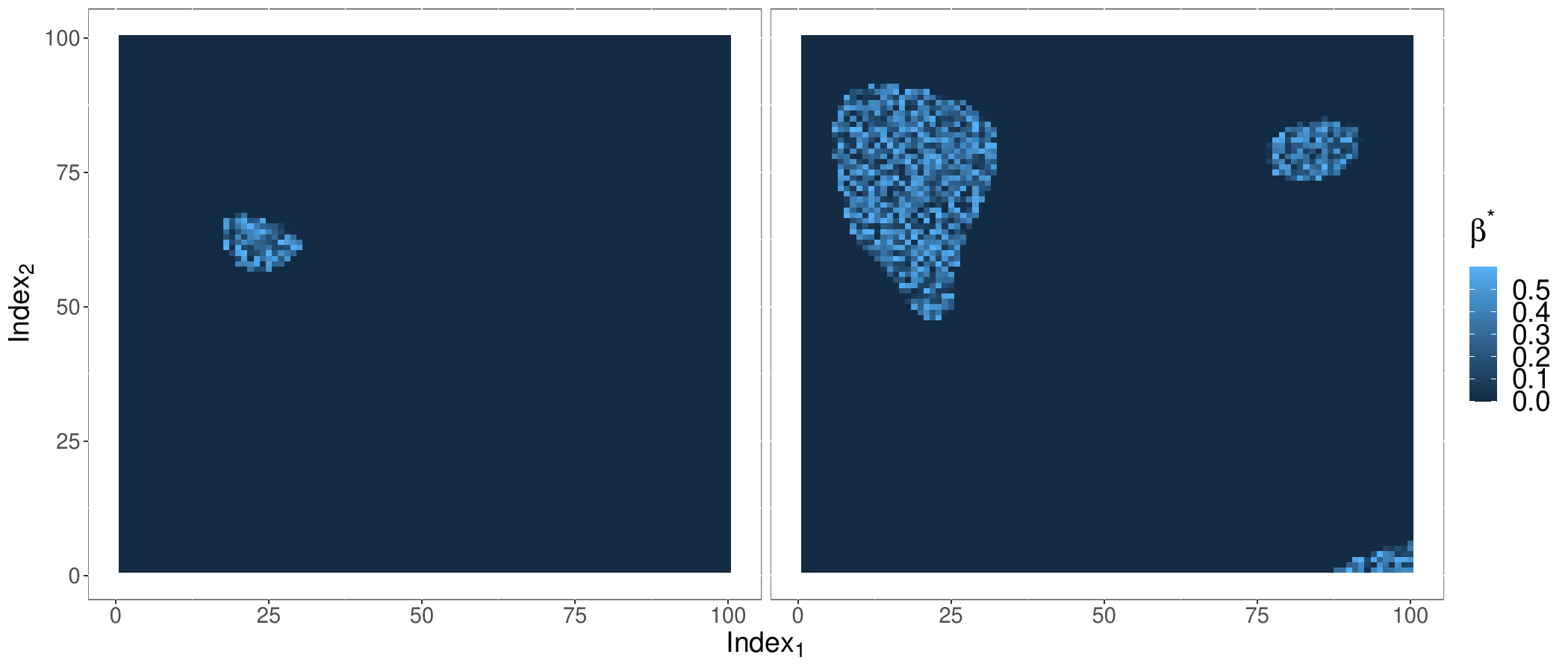} \\
 \includegraphics[scale=0.35]{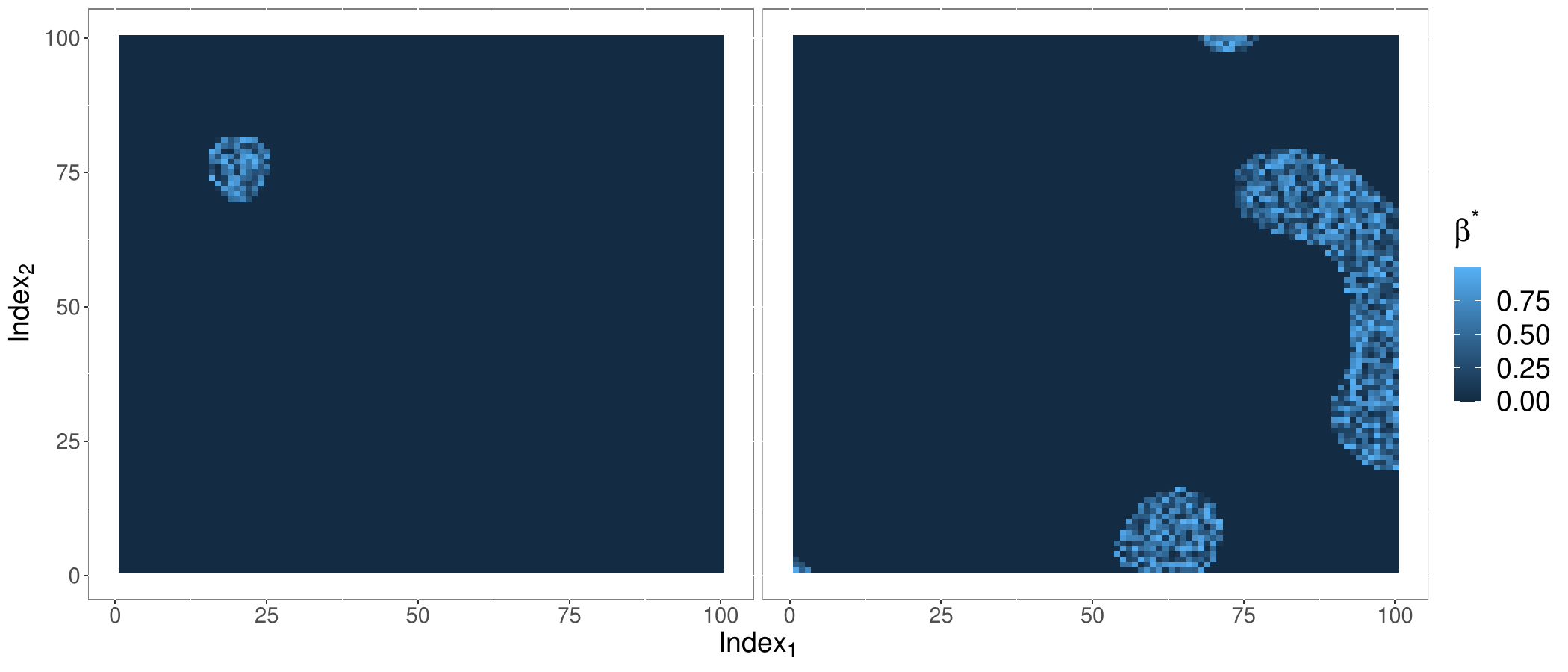} \\
 \includegraphics[scale=0.35]{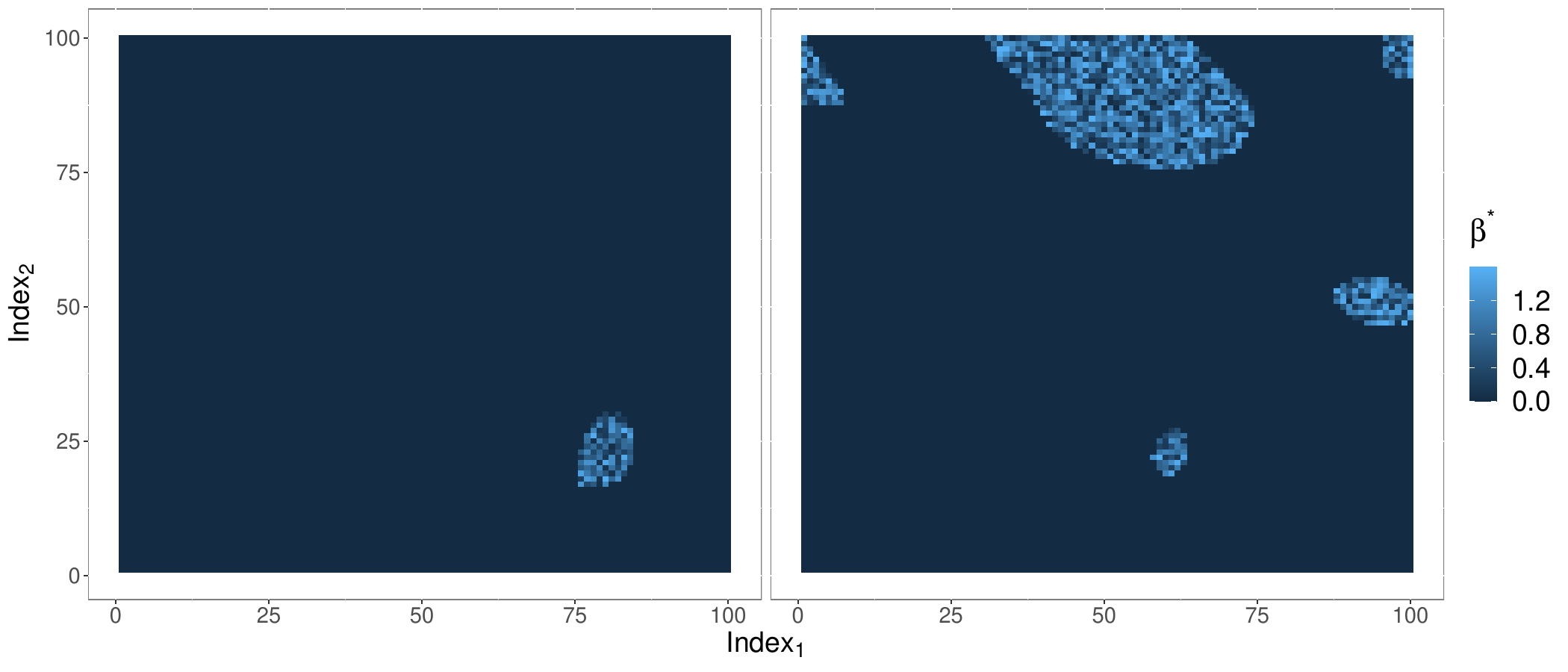} 
\end{tabular}
\caption{Examples of $\beast = \mbf{\gamma}\be$, for $M=10^4$, $M_1=100$ (left) and $M_1=10^3$ (right) where $\be \sim U(0,2\eta)$ with $\eta=0.3$ (top), $\eta=0.5$ (middle), and $\eta=0.8$ (bottom).\label{fig.signal.examp}}
\end{figure}

\begin{figure}[t]
\centering
\begin{tabular}{cc}
\includegraphics[scale=0.425]{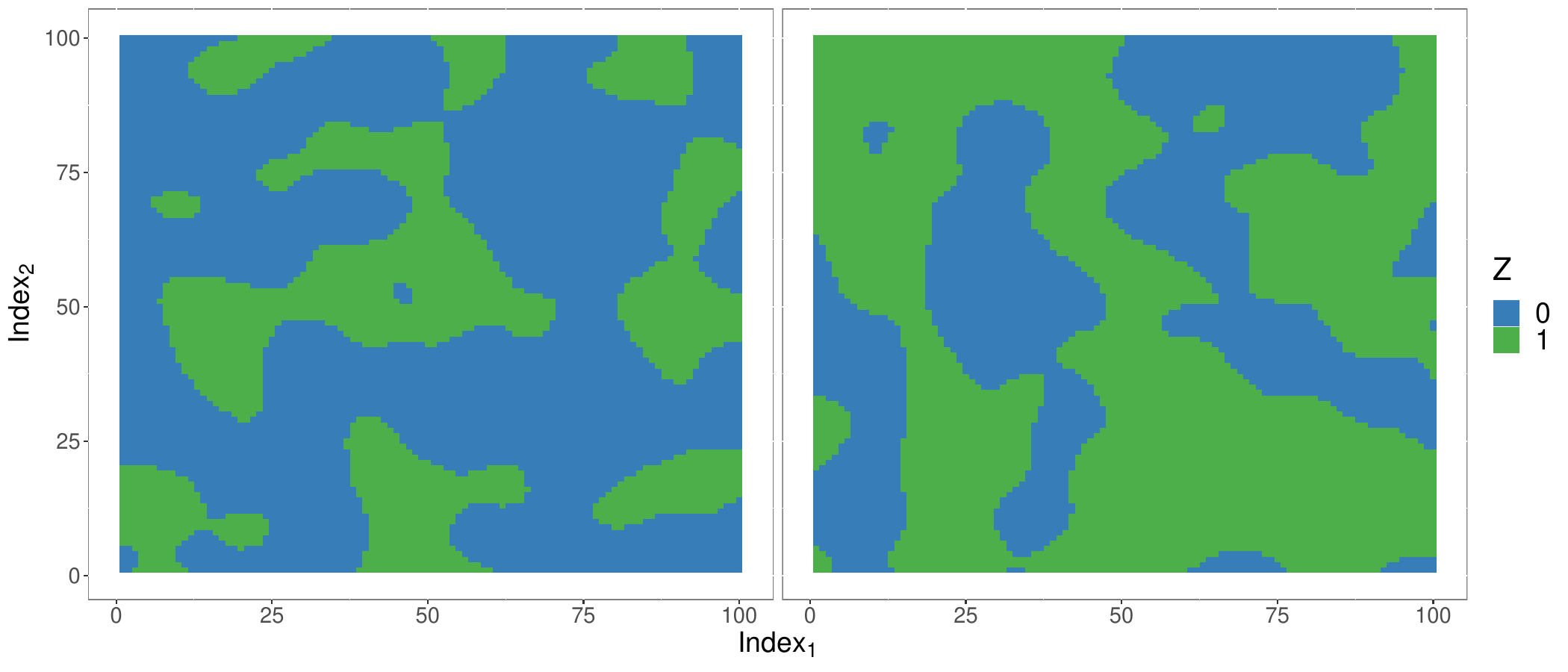} \\
\includegraphics[scale=0.425]{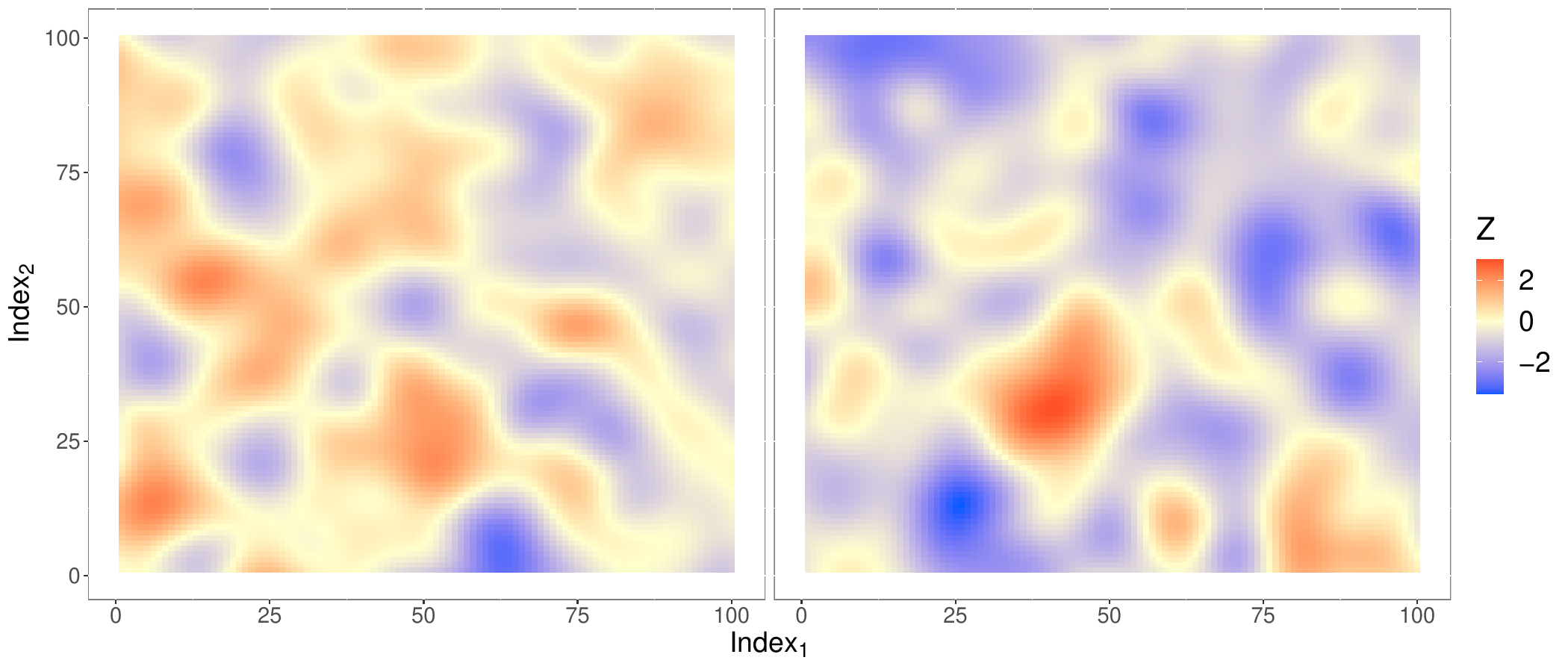} \\
\end{tabular}
\caption{Examples of binary (top) and continuous (bottom) $\mbf{X}$ data used in the simulation studies.\label{fig.data.examp}}
\end{figure}

Figures \ref{fig.signal.examp} and \ref{fig.data.examp} contain graphical examples of $\mbf{\gamma}\be$ and $\X$, respectively.  Each were generated randomly at each iteration.  Figure \ref{fig.signal.examp} show that $\mbf{\gamma}$ can have multiple spatial clusters of signals.

\section*{B. Data Access and additional results}

The data for Example 1 was downloaded from the Cancer Cell Line Encyclopedia online portal \citep[CCLE, \url{https://portals.broadinstitute.org/ccle},][]{Barretina2012}. The files were 
\begin{itemize}
\item \texttt{CCLE\_Expression\_Entrez\_2012\-09\-29.gct}, 
\item \texttt{CCLE\_NP24.2009\_Drug\_data\_2015.02.24.csv}, and 
\item \texttt{CCLE\_sample\_info\_file\_2012\-10\-18.txt}
\end{itemize}
Then, we followed Dondelinger et al. to use this dataset to model drug response \citep{Dondelinger2018}. We minimally modified the \texttt{R} script provided by Dondelinger et al. to process these data files \citep[see Section 8. Software Availability, in][]{Dondelinger2018}. The processed and ready-to-use data files and all \texttt{R} scripts to reproduce all results are available via \cite{McLZgo21}.


\begin{figure}[p]
\centering
\begin{tabular}{c}
\includegraphics[width=6in]{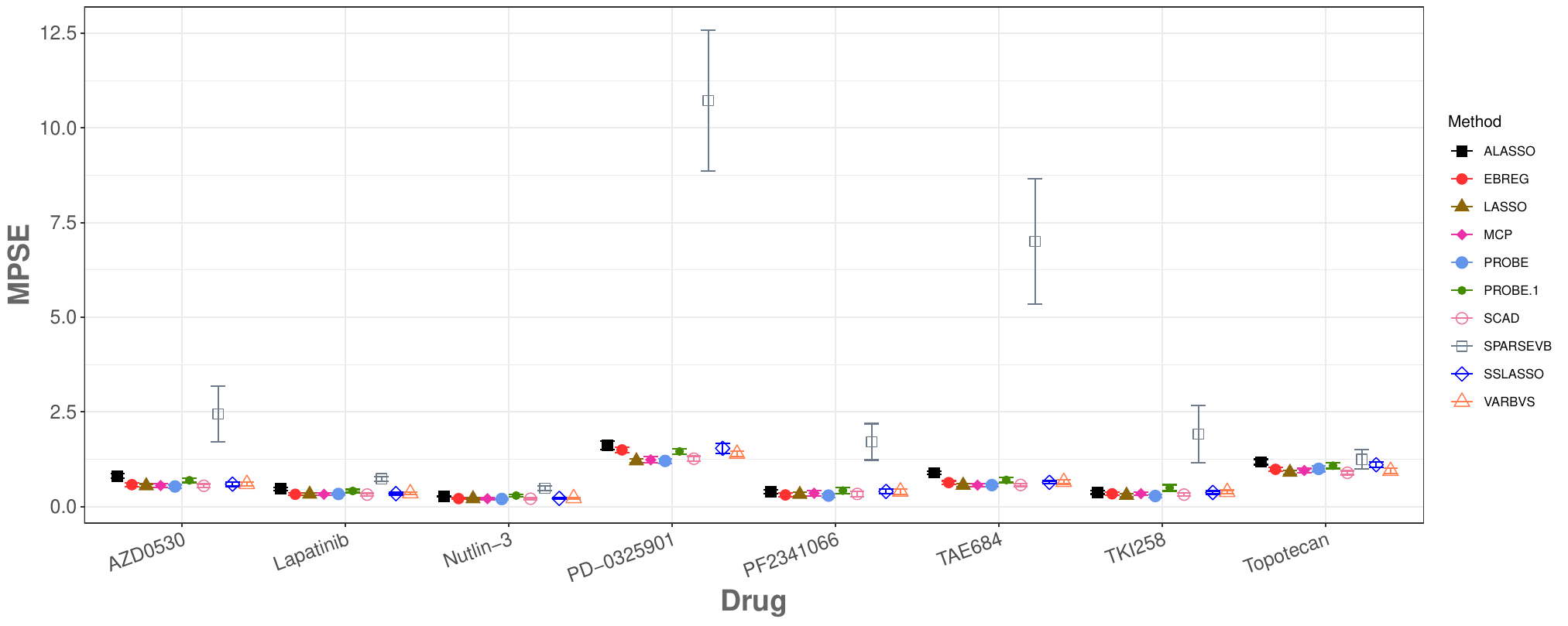} \\
\includegraphics[width=6in]{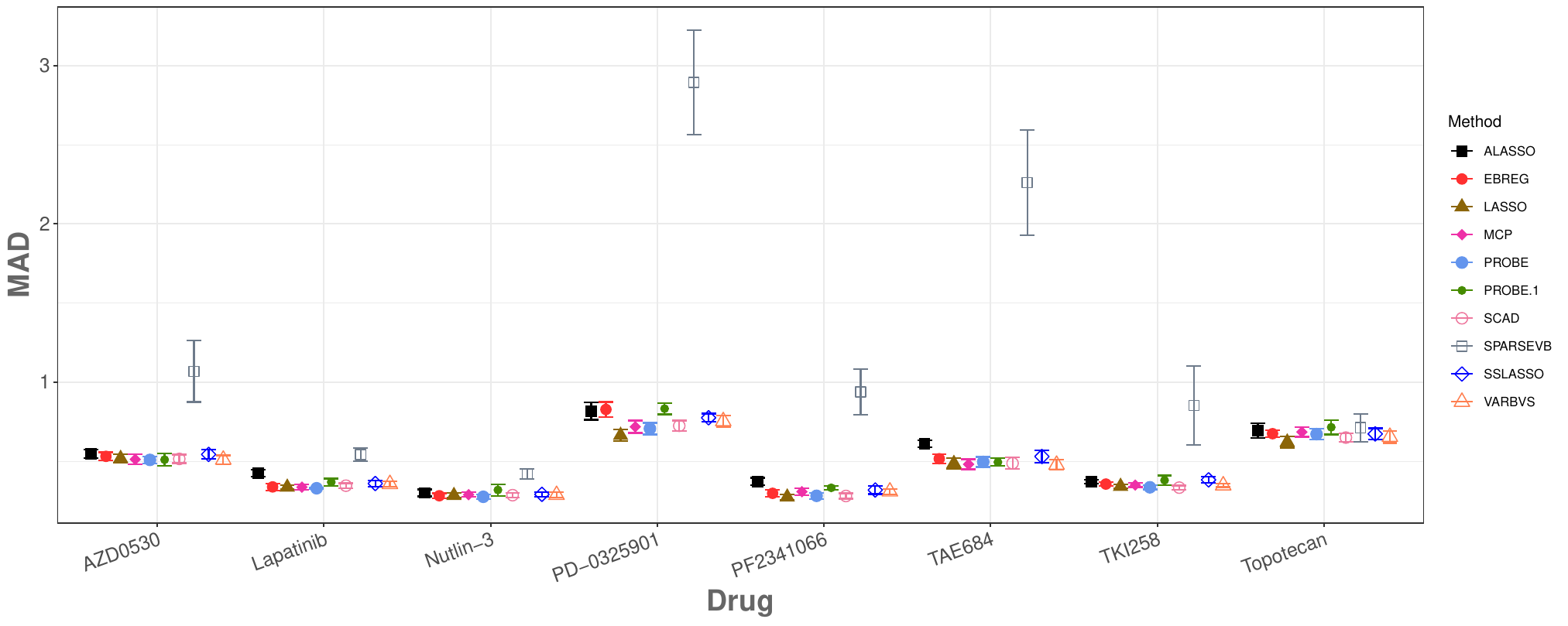} \\ 
\end{tabular}
\caption{For the eight anticancer drug response models (top) CV-MSPE including SPARSEVB, and (bottom) CV-MAD including SPARSEVB. Displayed is the overall MSPE/MAD $\pm$ SE (all based on 10-fold CV). \label{mse.mad.example1}}
\end{figure}

\end{document}